\documentclass[preprintnumbers]{revtex4}
\usepackage{amsfonts}
\usepackage{amsmath}
\usepackage{hyperref}
\usepackage{amssymb}
\usepackage[english]{babel}
\usepackage{graphicx}
\usepackage{epsfig}
\usepackage{bm}
\usepackage{longtable}
\usepackage{verbatim}
\usepackage{longtable}

\newcommand{\mathsym}[1]{{}}

\input{tcilatex}
\begin{document}

\title{Precision measurements constraints on the number of Higgs doublets}
\author{A. E. C\'arcamo Hern\'andez${}^{}$}
\email{antonio.carcamo@usm.cl}
\author{Sergey Kovalenko${}^{}$}
\email{sergey.kovalenko@usm.cl}
\author{Iv\'an Schmidt${}^{}$}
\email{ivan.schmidt@usm.cl}
\affiliation{${}^{}$Universidad T\'ecnica Federico Santa Mar\'{\i}a\\
and\\
Centro Cient\'{\i}fico-Tecnol\'ogico de Valpara\'{\i}so\\
Casilla 110-V, Valpara\'{\i}so, Chile}
\date{\today }

\begin{abstract}
We consider an extension of the Standard Model with an arbitrary number $N$
of Higgs doublets (NHDM), and calculate their contribution to the oblique
parameters $S$ and $T$. We examine the possible limitations on $N$ from 
precision measurements of these parameters. In view of the complexity of the
general case of NHDM, we analyze several benchmark scenarios for the
Higgs mass spectrum, identifying the lightest CP-even Higgs with the
Higgs-like particle recently observed at the LHC with the mass of $\sim 125$
GeV. The rest of the Higgses are put above the mass scale of $\sim 600$ GeV, below which the LHC experiments do
not detect any Higgs-like signals except for the former famous one. We
show that, in a scenario, with all the heavy Higgses degenerate at any scale,
there are no limitations on the number $N$ of the Higgs doublets. However,
upper limits appear for certain not completely degenerate configurations of
the heavy Higgses.

\end{abstract}

\maketitle

\section{Introduction}

The recent discovery  of the $\sim$125 GeV scalar particle at the Large Hadron Collider (LHC) 
\cite{:2012gk,:2012gu} perfectly fills the vacancy of the Higgs boson necessary for 
the completion of the Standard Model (SM) at the Fermi scale. Surprisingly, the SM with the Higgs boson in this mass range becomes formally self-consistent up to the Planck scale. In the absence of any signal of physics beyond the SM, this fact drastically strengthens the position of this model as the theoretical basis of particle physics. 

\quad Although the new observed scalar state has so far all the properties expected of the SM
Higgs boson, it is still possible that it could be a light scalar in a multi-Higgs extension of the SM, or a light supersymmetric
Higgs boson, or a Higgs boson coming from a strongly interacting dynamics,
where the theory becomes nonperturbative above the Fermi scale and the
breaking is achieved through some condensate. Now the priority of the LHC
experiments will be to measure precisely the couplings of the observed scalar 
to fermions and gauge bosons, and to establish its quantum numbers in order
to identify it with one of these or some other options.  On the other hand, searches for new particles beyond the SM are an essential task of the LHC experiments \cite{PDG,Brooijmans:2014eja,Carena:2013qia,Bechtle:2013wla,Heinemeyer:2014uoa,Dev:2014yca,Carena:2014nza,Bhattacharyya:2014oka}.

\quad In this paper, we consider a multi-Higgs extension of the SM, with an arbitrary number  $N$ of the Higgs electroweak doublets. Our goal is to study possible bounds on the number of Higgs doublets from the precision measurements of the oblique $T$ and $S$ parameters.

\quad We assume that the $N$ Higgs $SU(2)$
doublets are identical, with hypercharge equal to $1$. Some features such as
the relation between the mass and gauge eigenstates in the scalar sector and
the relation of the Higgs vacuum expectation values with the symmetry breaking scale 
$v\approx 246 $ GeV presented in the two Higgs doublet model are still fulfilled when the number of Higgs doublets is increased \cite{2N-model}.

\quad The paper is organized as follows. In Sec. \ref{model}, we
briefly describe the theoretical structure of the $N$ Higgs doublet model  (NHDM). In Sec. \ref{TandSparameters}, we compute the one-loop contribution to the $T$ and $S$ parameters in the NHDM. The bounds on the number of Higgs doublets coming from $T$ and $S$ parameter
constraints at $95\%$C.L. are computed in Sec. \ref{T-S-exp bounds}. In
Sec. \ref{Conclusions}, we summarize our results.

\section{The Model}

\label{model} We consider an extension of the SM with $N$ copies of the complex $%
SU(2)_{L}$ weak doublet scalar Higgs fields with  hypercharge $Y=1$ (NHDM). The model scalar potential, invariant with respect to the SM gauge group, is 
\begin{equation}  \label{V-scalar}
V=\frac{1}{2}\sum_{i,j=1}^{N}\mu _{ij}^{2}\Phi _{i}^{\dagger }\Phi _{j}+ 
\frac{1}{4}\sum_{i,j,k,l=1}^{N}\lambda _{ij,kl}\left( \Phi _{i}^{\dagger
}\Phi _{j}\right) \left( \Phi _{k}^{\dagger }\Phi _{l}\right)
+\sum_{i,j,k,l=1}^{N}\sigma _{ij,kl}\left( \Phi _{i}\tau^{2} \Phi
_{j}\right) \left( \Phi _{k} \tau^{2} \Phi _{l}\right)^{\dagger} .
\end{equation}
where $\tau ^{2}$ is a Pauli matrix in the $SU(2)_{L}$ space and 
\begin{equation}
\sigma _{ij,kl}=-\sigma _{ji,kl}=-\sigma _{ij,lk},\hspace{2cm}
\end{equation}
For simplicity, we assume all the parameters in the scalar potential to be real. Then
the Hermiticity of the scalar potential (\ref{V-scalar}) implies 
\begin{eqnarray}  \label{Prop-2}
\lambda _{ij,kl}=\lambda _{ji,lk},\hspace{5mm} \sigma_{ij,kl}=\sigma
_{ji,lk}, \hspace{5mm} \mu _{ij}=\mu _{ji},\hspace{2cm}\hspace{2cm}
\end{eqnarray}
The minimum of the scalar potential is parametrized by N vacuum expectation
values 
\begin{equation}
\left\langle \Phi _{l}\right\rangle =\left( 
\begin{array}{c}
0 \\ 
\frac{v_{l}}{\sqrt{2}}%
\end{array}%
\right) ,\hspace{2cm}\hspace{2cm}l=1,2,\cdots ,N.
\end{equation}%
We decompose the Higgs fields around this minimum as 
\begin{equation}
\Phi _{l}=\left( 
\begin{array}{c}
\phi _{l}^{+} \\ 
\frac{1}{\sqrt{2}}\left( v_{l}+\rho _{l}+i\eta _{l}\right)%
\end{array}%
\right) =\left( 
\begin{array}{c}
\frac{1}{\sqrt{2}}\left( \omega _{l}+i\xi _{l}\right) \\ 
\frac{1}{\sqrt{2}}\left( v_{l}+\rho _{l}+i\eta _{l}\right)%
\end{array}%
\right)  \label{doublets}
\end{equation}%
where 
\begin{equation}
\left\langle \rho _{l}\right\rangle =\left\langle \eta _{l}\right\rangle
=\left\langle \omega _{l}\right\rangle =\left\langle \xi _{l}\right\rangle
=0,\hspace{2cm}\hspace{2cm}l=1,2,\cdots ,N.
\end{equation}%
Then the covariant derivative acting on the Higgs doublets takes the form 
\begin{eqnarray}
D_{\mu }\Phi _{l} &=&\partial _{\mu }\Phi _{l}-\frac{i}{2}gW_{\mu }^{a}\tau
^{a}\Phi _{l}-\frac{i}{2}g^{\prime }Y_{l}B_{\mu }\Phi _{l}  \notag \\
&=&\left( 
\begin{array}{c}
\frac{1}{\sqrt{2}}\partial _{\mu }\omega _{l}+\frac{1}{2\sqrt{2}}\left[
gW_{\mu }^{1}\eta _{l}-gW_{\mu }^{2}\left( v_{l}+\rho _{l}\right) +\left(
gW_{\mu }^{3}+g^{\prime }Y_{l}B_{\mu }\right) \xi _{l}\right] \\ 
\frac{1}{\sqrt{2}}\partial _{\mu }\rho _{l}+\frac{1}{2\sqrt{2}}\left(
gW_{\mu }^{1}\xi _{l}+gW_{\mu }^{2}\omega _{l}-gW_{\mu }^{3}\eta
_{l}+g^{\prime }Y_{l}B_{\mu }\eta _{l}\right)%
\end{array}%
\right)  \notag \\
&&+\frac{i}{2\sqrt{2}}\left( 
\begin{array}{c}
2\partial _{\mu }\xi -\left( gW_{\mu }^{3}+g^{\prime }Y_{l}B_{\mu }\right)
\omega _{l}-\left[ gW_{\mu }^{1}\left( v_{l}+\rho _{l}\right) +gW_{\mu
}^{2}\eta _{l}\right] \\ 
2\partial _{\mu }\eta _{l}-gW_{\mu }^{1}\omega _{l}+gW_{\mu }^{2}\xi
_{l}-\left( -gW_{\mu }^{3}+g^{\prime }Y_{l}B_{\mu }\right) \left( v_{l}+\rho
_{l}\right)%
\end{array}%
\right) ,
\end{eqnarray}%
where the $\tau ^{a}$ are the ordinary $SU(2)_{L}$ Pauli matrices and $%
Y_{l}=1$.\newline
The NHDM scalar-gauge boson interactions are given by 
\begin{eqnarray}
\dsum\limits_{l=1}^{N}\left( D_{\mu }\Phi _{l}\right) \left( D^{\mu }\Phi
_{l}\right) ^{\dag } &=&\frac{1}{8}\dsum\limits_{l=1}^{N}\left\{ 2\partial
_{\mu }\omega _{l}+\left[ gW_{\mu }^{1}\eta _{l}-gW_{\mu }^{2}\left(
v_{l}+\rho _{l}\right) +\left( gW_{\mu }^{3}+g^{\prime }Y_{l}B_{\mu }\right)
\xi _{l}\right] \right\} ^{2}  \notag \\
&&+\frac{1}{8}\dsum\limits_{l=1}^{N}\left\{ 2\partial _{\mu }\xi _{l}-\left(
gW_{\mu }^{3}+g^{\prime }Y_{l}B_{\mu }\right) \omega _{l}-\left[ gW_{\mu
}^{1}\left( v_{l}+\rho _{l}\right) +gW_{\mu }^{2}\eta _{l}\right] \right\}
^{2}  \notag \\
&&+\frac{1}{8}\dsum\limits_{l=1}^{N}\left\{ 2\partial _{\mu }\rho _{l}+\left[
gW_{\mu }^{1}\xi _{l}+gW_{\mu }^{2}\omega _{l}-\left( gW_{\mu
}^{3}-g^{\prime }Y_{l}B_{\mu }\right) \right] \eta _{l}\right\} ^{2}  \notag
\\
&&+\frac{1}{8}\dsum\limits_{l=1}^{N}\left\{ 2\partial _{\mu }\eta
_{l}-gW_{\mu }^{1}\omega _{l}+gW_{\mu }^{2}\xi _{l}-\left( -gW_{\mu
}^{3}+g^{\prime }Y_{l}B_{\mu }\right) \left( v_{l}+\rho _{l}\right) \right\}
^{2}.  \label{GBSInt}
\end{eqnarray}%

The connection between the interaction and mass scalar eigenstates is
explained in what follows. The charged scalar fields of Eq. (\ref{doublets})
are linear combinations of the charged Goldstone bosons and the charged physical
scalars. The imaginary parts of the neutral component of the scalar doublets
of Eq. (\ref{doublets}) are linear combinations of the neutral Goldstone
bosons and of the CP-odd neutral scalar fields. The real parts of the neutral
component of the scalar doublets of Eq. (\ref{doublets}) are linear
combinations of the CP-odd neutral scalar fields. Within this framework we
consider a scenario where the interaction and mass eigenstates are related
in the way analogous to the two Higgs doublet model (2HDM) \cite{2N-model} 

\begin{equation}  \label{Neutral}
\rho _{l}=\sum_{j=1}^{N}R_{lj}H_{j}^{0},\hspace{1.5cm}\hspace{1.5cm}\eta
_{l}=Q_{l1}\pi ^{0}+\sum_{j=2}^{N}Q_{lj}A_{j-1}^{0},\hspace{2cm}l=1,2,\cdots
,N.
\end{equation}%
\begin{equation}  \label{Comp-Charged}
\omega _{l}=Q_{l1}\pi ^{1}+\sum_{j=2}^{N}Q_{lj}H_{j-1}^{1},\hspace{1.5cm}%
\hspace{1.5cm}\xi _{l}=Q_{l1}\pi ^{2}+\sum_{j=2}^{N}Q_{lj}H_{j-1}^{2},%
\hspace{2cm}l=1,2,\cdots ,N.
\end{equation}%
where:%
\begin{equation}  \label{vev}
v_{l}=vQ_{l1},\hspace{1.2cm}l=1,2,\cdots ,N,\hspace{1.2cm}%
v^{2}=\sum_{i=1}^{N}v_{l}^{2},\hspace{1.2cm}\sum_{l=1}^{N}R_{li}R_{lj}=%
\delta _{ij},\hspace{1.2cm}\sum_{l=1}^{N}Q_{li}Q_{lj}=\delta _{ij}.
\end{equation}
Here $v\approx 246$ GeV is the conventional electroweak symmetry breaking
scale. The fields $H_{i}^{0}$ ($i=1,2,\cdots ,N$) and $A_{j}^{0}$ ($%
j=1,2,\cdots ,N-1$) are the CP-even and CP-odd neutral Higgs bosons,
respectively. Similarly to the $W^{\pm}$ gauge bosons which are defined in terms of $W^1$ and $W^2$, the charged Higgs and Goldstone bosons are related to the component fields in (\ref{Comp-Charged}) as 
\begin{eqnarray}  \label{Charged}
H^{\pm}_{j} = \frac{H^{1}_{j} \mp i H^{2}_{j}}{\sqrt{2}}, \ \ \ \ \pi^{\pm}
= \frac{\pi^{1} \mp i \pi^{2}}{\sqrt{2}}
\end{eqnarray}

Thus we assumed the following: 

\begin{enumerate}
\item The rotation matrix $Q$, which relates the neutral Goldstone boson $%
\pi^{0}$ and the CP odd neutral Higgses $A^{0}_{j}$ with the interaction
eigenstate scalars $\eta_l$ ($l=1,2,\cdots ,N$) in Eq. (\ref{Neutral}), is
the same as the one that relates the components of the charged Goldstone bosons $%
\pi^{1,2}$ and Higgses $H^{1,2}_{j}$ with the corresponding interaction
eigenstates $\omega_{l}, \xi_{l}$, ($l=1,2,\cdots ,N$) in Eqs. (\ref{Comp-Charged}), (\ref{Charged}). 

\item The vacuum expectation values of $N$ Higgs fields $v_{l}$ ($%
l=1,2,\cdots ,N$) are related to the common symmetry breaking scale $%
v\approx 246$ GeV through the first relation in Eq. (\ref{vev}).
\end{enumerate}

Both assumptions are generalizations of the corresponding relations of the
2HDM \cite{2N-model}. In the case of NHDM, these relations are not true
everywhere in the parametric space but only in a certain part of it. Adopting the above assumptions, we limit
ourselves to a region in the parametric space of the NHDM, which is motivated
(hinted) by the 2HDM.

\section{One-loop contribution to the $T$ and $S$ parameters.}

\label{TandSparameters} In this section we calculate one-loop contributions to
the oblique parameters $T$ and $S$ defined as \cite{Peskin:1991sw,Peskin:1991sw2,epsilon-approach,epsilon-approach2,Barbieri:2004,Barbieri-book}%
: 
\begin{equation}
T=\frac{\Pi _{33}\left( q^{2}\right) -\Pi _{11}\left( q^{2}\right) }{\alpha
_{EM}(M_{Z})M_{W}^{2}}\biggl|_{q^{2}=0},\ \ \ \ \ \ \ \ \ \ \ S=\frac{2\sin 2%
{\theta }_{W}}{\alpha _{EM}(M_{Z})}\frac{d\Pi _{30}\left( q^{2}\right) }{%
dq^{2}}\biggl|_{q^{2}=0}.  \label{T-S-definition}
\end{equation}%
Here $\Pi _{11}\left( 0\right) $, $\Pi _{33}\left( 0\right) $, and $\Pi
_{30}\left( q^{2}\right) $ are the vacuum polarization amplitudes with $%
\{W_{\mu }^{1},W_{\mu }^{1}\}$, $\{W_{\mu }^{3},W_{\mu }^{3}\}$ and $\{W_{\mu }^{3},B_{\mu }\}$ external gauge bosons, respectively, where $q$ is
their momentum. Let us note that, in the aforementioned definitions of the oblique $T$ and $S$ parameters, it is assumed that the new physics is not light compared to $M_W$ and $M_Z$.

\subsection{$T$ parameter}

\label{subsection T-parameter}

The interaction Lagrangian, relevant for the computation of one-loop
contributions to the $T$ parameter in Eq. (\ref{T-S-definition}),  is 
\begin{eqnarray}  \label{T-Lagrangian}
\tciLaplace _{int}^{\left( T\right) } &=&\frac{gg^{\prime }v}{2}%
\pi^{1}W^{1\mu }B_{\mu }+\frac{gg^{\prime }v}{2}
\sum_{i=1}^{N}P_{i1}H_{i}^{0}W^{3\mu }B_{\mu }+\frac{g}{2}\left(
\pi^{0}\partial _{\mu }\pi ^{1}-\pi ^{1}\partial _{\mu }\pi ^{0}\right)
W^{1\mu}+ \frac{g}{2}\sum_{i=1}^{N-1}\left( A_{i}^{0}\partial
_{\mu}H_{i}^{1}-H_{i}^{1}\partial _{\mu }A_{i}^{0}\right) W^{1\mu }  \notag
\\
&&+\frac{g}{2}\sum_{i=1}^{N}P_{i1}\left( \pi ^{2}\partial
_{\mu}H_{i}^{0}-H_{i}^{0}\partial _{\mu }\pi ^{2}\right) W^{1\mu }+\frac{g}{2%
} \sum_{i=1}^{N}\sum_{j=1}^{N-1}P_{i,j+1}\left( H_{j}^{2}\partial
_{\mu}H_{i}^{0}-H_{i}^{0}\partial _{\mu }H_{j}^{2}\right) W^{1\mu }  \notag
\\
&&+\frac{g}{2}\left( \pi ^{2}\partial _{\mu }\pi ^{1}-\pi ^{1}\partial
_{\mu}\pi ^{2}\right) W^{3\mu }+ \frac{g}{2}\sum_{i=1}^{N-1}\left(H_{i}^{2}%
\partial _{\mu }H_{i}^{1}-H_{i}^{1}\partial _{\mu }H_{i}^{2}\right) W^{3\mu
}+\frac{g}{2}\sum_{i=1}^{N}P_{i1}\left( H_{i}^{0}\partial _{\mu }\pi
^{0}-\pi ^{0}\partial _{\mu }H_{i}^{0}\right) W^{3\mu }  \notag \\
&&+\frac{g}{2}\sum_{i=1}^{N}\sum_{j=1}^{N-1}P_{i,j+1}\left(H_{i}^{0}\partial
_{\mu }A_{j}^{0}-A_{j}^{0}\partial _{\mu }H_{i}^{0}\right) W^{3\mu }.
\end{eqnarray}
where $P_{ij}$ is given by 
\begin{equation}  \label{P-par}
P_{ij}=\sum_{l=1}^{N}R_{li}Q_{lj}.
\end{equation}
By definition it satisfies the inequality 
\begin{eqnarray}  \label{P-lim}
0 \leq P_{ij}\leq 1
\end{eqnarray}

\begin{figure}[tbh]
\includegraphics[width=15cm,height=20cm,angle=0]{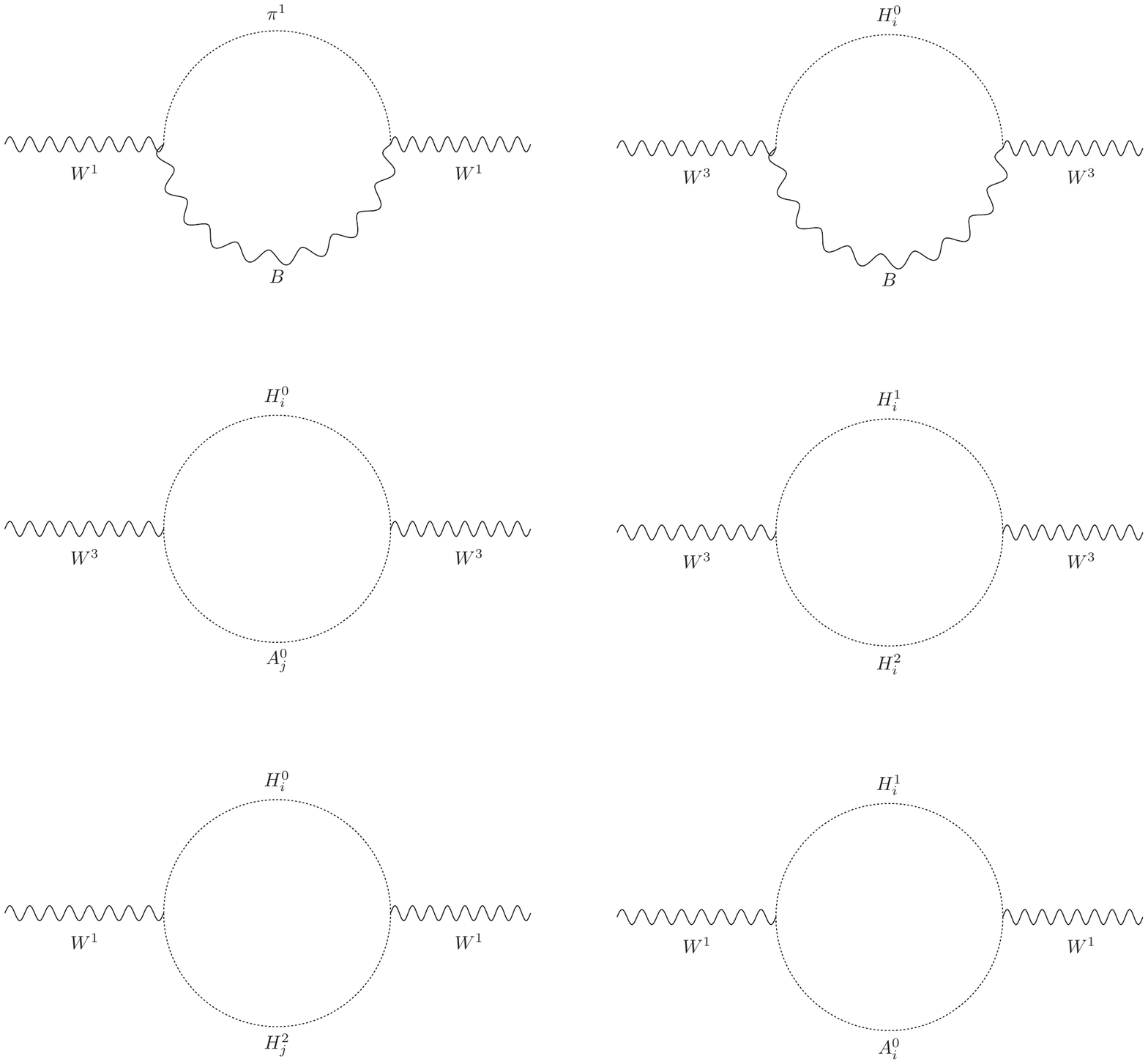} 
\vspace{-35mm} \vspace{5mm}
\caption{One-loop Feynman diagrams contributing to the $T$ parameter.}
\label{diag1}
\end{figure}

As seen from Eq. (\ref{T-Lagrangian}), the $T$ parameter 
(\ref{T-S-definition}) at one-loop level receives contributions from the diagrams shown in 
Fig.~\ref{diag1}. 

Their partial contributions, assuming the cutoff $\Lambda $ to be
much larger than the masses of the scalar particles, are 
\begin{equation}
T_{\left( \pi ^{1}B\right) }\simeq -\frac{3}{16\pi \cos ^{2}\theta _{W}}\ln
\left( \frac{\Lambda ^{2}}{m_{W}^{2}}\right) ,
\end{equation}%
\begin{equation}
\sum_{i=1}^{N}T_{\left( H_{i}^{0}B\right) }\simeq \frac{3}{16\pi \cos
^{2}\theta _{W}}\sum_{i=1}^{N}P_{i1}^{2}\ln \left( \frac{\Lambda ^{2}}{%
m_{H_{i}^{0}}^{2}}\right) ,
\end{equation}%
\begin{equation}
\sum_{i=1}^{N}\sum_{j=1}^{N-1}T_{\left( H_{i}^{0}A_{j}^{0}\right) }\simeq\frac{1}{%
16\alpha _{EM}(M_{Z})\pi ^{2}v^{2}}\sum_{i=1}^{N}%
\sum_{j=1}^{N-1}P_{i,j+1}^{2}F\left( \Lambda
^{2},m_{H_{i}^{0}}^{2},m_{A_{j}^{0}}^{2}\right) ,
\end{equation}%
\begin{equation}
\sum_{i=1}^{N-1}T_{\left( H_{i}^{1}H_{i}^{2}\right) }\simeq\frac{1}{16\alpha
_{EM}(M_{Z})\pi ^{2}v^{2}}\sum_{i=1}^{N-1}G\left( \Lambda ^{2},m_{H_{i}^{\pm
}}^{2}\right) ,
\end{equation}%
\begin{equation}
\sum_{i=1}^{N}\sum_{j=1}^{N-1}T_{\left( H_{i}^{0}H_{j}^{2}\right) }\simeq-\frac{1%
}{16\alpha _{EM}(M_{Z})\pi ^{2}v^{2}}\sum_{i=1}^{N}%
\sum_{j=1}^{N-1}P_{i,j+1}^{2}F\left( \Lambda
^{2},m_{H_{i}^{0}}^{2},m_{H_{j}^{\pm }}^{2}\right) ,
\end{equation}%
\begin{equation}
\sum_{i=1}^{N-1}T_{\left( H_{i}^{1}A_{i}^{0}\right) }\simeq-\frac{1}{16\alpha
_{EM}(M_{Z})\pi ^{2}v^{2}}\sum_{i=1}^{N-1}F\left( \Lambda ^{2},m_{H_{i}^{\pm
}}^{2},m_{A_{i}^{0}}^{2}\right).
\end{equation}
The subscripts in $T_{ab}$  denote the internal lines of the diagrams in Fig.~\ref{diag1}.
The functions $F\left( \Lambda ^{2},m_{1}^{2},m_{2}^{2}\right) $ and 
$G\left( \Lambda ^{2},m^{2}\right) $ are defined as 
\begin{equation}
F\left( \Lambda ^{2},m_{1}^{2},m_{2}^{2}\right)=\Lambda ^{2}-\frac{m_{1}^{4}}{m_{1}^{2}-m_{2}^{2}}\ln \left( \frac{\Lambda ^{2}}{m_{1}^{2}}\right) -\frac{m_{2}^{4}}{m_{2}^{2}-m_{1}^{2}}\ln \left( \frac{\Lambda^{2}}{m_{2}^{2}}\right) ,
\end{equation}%
\begin{equation}
G\left( \Lambda ^{2},m^{2}\right) =\lim_{m_{1},m_{2}\rightarrow m}F\left(
\Lambda ^{2},m_{1},m_{2}\right) =\allowbreak \Lambda ^{2}-2m^{2}\ln \left( \frac{\Lambda ^{2}}{m^{2}}\right) +m^2
\end{equation}%
Collecting all the contributions together, we find the one-loop contribution to
the $T$ parameter coming from the scalar sector of the NHDM: 
\begin{eqnarray}
T = \sum_{ab}T_{ab} &\simeq &-\frac{3}{16\pi \cos ^{2}\theta _{W}}\sum_{i=1}^{N}P_{i1}^{2}\ln
\left( \frac{m_{H_{i}^{0}}^{2}}{m_{W}^{2}}\right) +\frac{1}{16\alpha
_{EM}(M_{Z})\pi ^{2}v^{2}}\sum_{i=1}^{N-1}\left[ m_{H_{i}^{\pm
}}^{2}-h\left( m_{A_{i}^{0}}^{2},m_{H_{i}^{\pm }}^{2}\right) \right]   \notag
\\
&&+\frac{1}{16\alpha _{EM}(M_{Z})\pi ^{2}v^{2}}\sum_{i=1}^{N}%
\sum_{j=1}^{N-1}P_{i,j+1}^{2}\left[ h\left(
m_{H_{i}^{0}}^{2},m_{A_{j}^{0}}^{2}\right) -h\left(
m_{H_{i}^{0}}^{2},m_{H_{j}^{\pm }}^{2}\right) \right]   \notag \\
&=&-\frac{3}{16\pi \cos ^{2}\theta _{W}}\ln \left( \frac{m_{h}^{2}}{m_{W}^{2}%
}\right) +\frac{3\left( 1-P_{11}^{2}\right) }{16\pi \cos ^{2}\theta _{W}}\ln \left( \frac{m_{H^{0}}^{2}}{%
m_{h}^{2}}\right)   \notag \\
&&+\frac{N-1}{16\alpha _{EM}(M_{Z})\pi ^{2}v^{2}}\left[ m_{H_{i}^{\pm
}}^{2}-h\left( m_{A_{i}^{0}}^{2},m_{H_{i}^{\pm }}^{2}\right) \right]   \notag
\\
&&+\frac{1}{16\alpha _{EM}(M_{Z})\pi ^{2}v^{2}}\sum_{i=2}^{N}%
\sum_{j=1}^{N-1}P_{i,j+1}^{2}\left[ h\left(
m_{H_{i}^{0}}^{2},m_{A_{j}^{0}}^{2}\right) -h\left(
m_{H_{i}^{0}}^{2},m_{H_{j}^{\pm }}^{2}\right) \right] ,  \label{Th}
\end{eqnarray}%
where we identified the lightest CP-even Higgs $H_{1}^{0}=h$ with the LHC
Higgs-like particle with the mass $m_{h}=125$ GeV.

The function $h\left( m_{1}^{2},m_{2}^{2}\right) $ is given by: 
\begin{equation}
h\left( m_{1}^{2},m_{2}^{2}\right) =\frac{m_{1}^{2}m_{2}^{2}}{%
m_{1}^{2}-m_{2}^{2}}\ln \left( \frac{m_{1}^{2}}{m_{2}^{2}}\right) ,\hspace{%
1.5cm}\hspace{1.5cm}\lim_{m_{2}\rightarrow m_{1}}h\left(
m_{1}^{2},m_{2}^{2}\right) =m_{1}^{2}.
\end{equation}%

We can split the $T$ parameter as $T=T_{SM}+\Delta T$, where $T_{SM} $ is
the contribution from the SM, while $\Delta T$ contain all the
contributions involving the heavy scalars: 
\begin{equation}
T_{SM}=-\frac{3}{16\pi \cos ^{2}\theta _{W}}\ln \left( \frac{m_{h}^{2}}{%
m_{W}^{2}}\right) ,  \label{Tt}
\end{equation}%
\begin{eqnarray}
\label{Delta-T-1}
\Delta T &\simeq &-\frac{3}{16\pi \cos ^{2}\theta _{W}}%
\sum_{i=2}^{N}P_{i1}^{2}\ln \left( \frac{m_{H_{i}^{0}}^{2}}{m_{h}^{2}}%
\right) +\frac{1}{16\pi ^{2}v^{2}\alpha _{EM}(M_{Z})}\sum_{i=1}^{N-1}\left[
m_{H_{i}^{\pm }}^{2}-h\left( m_{A_{i}^{0}}^{2},m_{H_{i}^{\pm }}^{2}\right) %
\right] \\
\nonumber
&&+\frac{1}{16\pi ^{2}v^{2}\alpha _{EM}(M_{Z})}\sum_{i=1}^{N}%
\sum_{j=1}^{N-1}P_{i,j+1}^{2}\left[ h\left(
m_{H_{i}^{0}}^{2},m_{A_{j}^{0}}^{2}\right) -h\left(
m_{H_{i}^{0}}^{2},m_{H_{j}^{\pm }}^{2}\right) \right] .
\end{eqnarray}

\subsection{$S$ parameter}

\label{subsection S-parameter}

The interaction Lagrangian relevant for the computation of the one-loop
contribution to the $S$ parameter in Eq. (\ref{T-S-definition}) is 
\begin{eqnarray}
\tciLaplace _{int}^{\left( S\right) } &=&\frac{g}{2}\left( \pi ^{2}\partial
_{\mu }\pi ^{1}-\pi ^{1}\partial _{\mu }\pi ^{2}\right) W^{3\mu }+\frac{g}{2}%
\sum_{i=1}^{N-1}\left( H_{i}^{2}\partial _{\mu }H_{i}^{1}-H_{i}^{1}\partial
_{\mu }H_{i}^{2}\right) W^{3\mu }  \notag \\
&&+\frac{g^{\prime }}{2}\left( \pi ^{2}\partial _{\mu }\pi ^{1}-\pi
^{1}\partial _{\mu }\pi ^{2}\right) B^{\mu }+\frac{g^{\prime }}{2}%
\sum_{i=1}^{N-1}\left( H_{i}^{2}\partial _{\mu }H_{i}^{1}-H_{i}^{1}\partial
_{\mu }H_{i}^{2}\right) B^{\mu }  \notag \\
&&+\frac{g}{2}\sum_{i=1}^{N}P_{i1}\left( H_{i}^{0}\partial _{\mu }\pi
^{0}-\pi ^{0}\partial _{\mu }H_{i}^{0}\right) W^{3\mu }+\frac{g}{2}%
\sum_{i=1}^{N}\sum_{j=1}^{N-1}P_{i,j+1}\left( H_{i}^{0}\partial _{\mu
}A_{j}^{0}-A_{j}^{0}\partial _{\mu }H_{i}^{0}\right) W^{3\mu }  \notag \\
&&-\frac{g^{\prime }}{2}\sum_{i=1}^{N}P_{i1}\left( H_{i}^{0}\partial _{\mu
}\pi ^{0}-\pi ^{0}\partial _{\mu }H_{i}^{0}\right) B^{\mu }-\frac{g^{\prime }%
}{2}\sum_{i=1}^{N}\sum_{j=1}^{N-1}P_{i,j+1}\left( H_{i}^{0}\partial _{\mu
}A_{j}^{0}-A_{j}^{0}\partial _{\mu }H_{i}^{0}\right) B^{\mu }.
\end{eqnarray}

\begin{figure}[tbh]
\centering
\vspace{-30mm} \vspace{0mm}%
\includegraphics[width=15cm,height=20cm,angle=0]{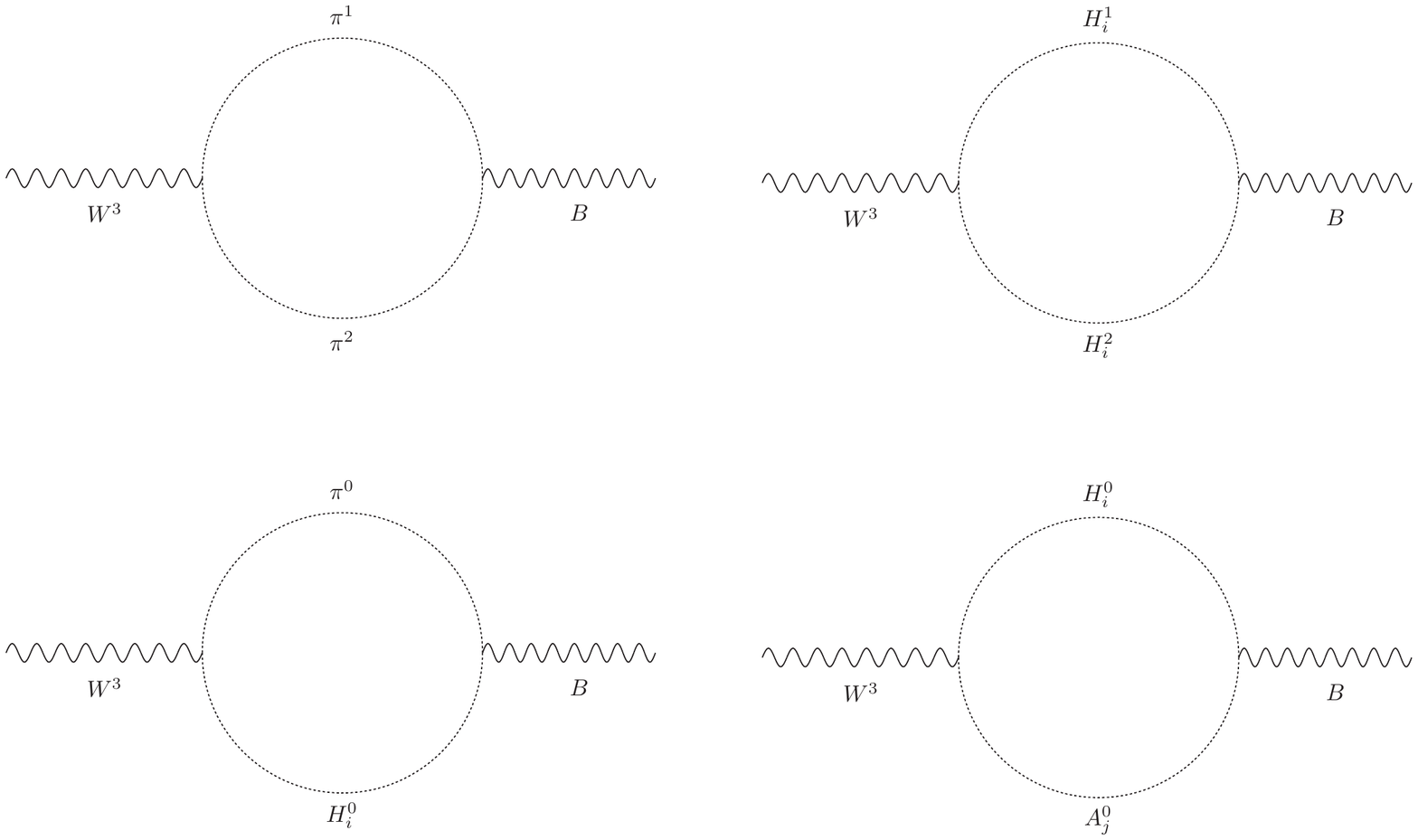} 
\vspace{-60mm} \vspace{10mm}
\caption{One-loop Feynman diagrams contributing to the $S$ parameter.}
\label{diag-2}
\end{figure}

As follows from this Lagrangian and the definition (\ref{T-S-definition}),
the $S$ parameter at one-loop level receives contributions from the diagrams shown in 
Fig.~\ref{diag-2}. 

Their partial contributions, assuming the cutoff $\Lambda $ to be
much larger than the masses of the scalar particles, are 
\begin{equation}
S_{\left( \pi ^{1}\pi ^{2}\right) }\simeq \frac{1}{12\pi }\ln \left( \frac{%
\Lambda ^{2}}{m_{W}^{2}}\right) ,
\end{equation}%
\begin{equation}
\sum_{i=1}^{N-1}S_{\left( H_{i}^{1}H_{i}^{2}\right) }\simeq \frac{1}{12\pi }%
\sum_{i=1}^{N-1}\ln \left( \frac{\Lambda ^{2}}{m_{H_{i}^{\pm }}^{2}}\right) ,
\end{equation}%
\begin{equation}
\sum_{i=1}^{N}S_{\left( H_{i}^{0}\pi ^{0}\right) }\simeq -\frac{1}{12\pi }%
\sum_{i=1}^{N}P_{i1}^{2}\ln \left( \frac{\Lambda ^{2}}{m_{H_{i}^{0}}^{2}}%
\right) ,
\end{equation}%
\begin{eqnarray}
\sum_{i=1}^{N}\sum_{j=1}^{N-1}S_{\left( H_{i}^{0}A_{j}^{0}\right) } &\simeq
&-\frac{1}{12\pi }\sum_{i=1}^{N}\sum_{j=1}^{N-1}\frac{P_{i,j+1}^{2}}{\left(
m_{A_{j}^{0}}^{2}-m_{H_{i}^{0}}^{2}\right) {}^{3}}\left\{ m_{A_{j}^{0}}^{6}%
\left[ \ln \left( \frac{\Lambda ^{2}}{m_{A_{j}^{0}}^{2}}\right) +\frac{5}{6}%
\right] -m_{H_{i}^{0}}^{6}\left[ \ln \left( \frac{\Lambda ^{2}}{%
m_{H_{i}^{0}}^{2}}\right) +\frac{5}{6}\right] \right.   \notag \\
&&+\left. 3m_{H_{i}^{0}}^{2}m_{A_{j}^{0}}^{2}\left[ m_{H_{i}^{0}}^{2}\left[
\ln \left( \frac{\Lambda ^{2}}{m_{H_{i}^{0}}^{2}}\right) +\frac{3}{2}\right]
-m_{A_{j}^{0}}^{2}\left[ \ln \left( \frac{\Lambda ^{2}}{m_{A_{j}^{0}}^{2}}%
\right) +\frac{3}{2}\right] \right] \right\} .
\end{eqnarray}
As before, the subscripts in $S_{ab}$ denote the internal lines of the diagrams in 
Fig.~\ref{diag-2}.
Then, the 1-loop Higgs contribution to the $S$ parameter in the NHDM is 
\begin{eqnarray}
S = \sum_{ab}S_{ab}&\simeq &\frac{1}{12\pi }\left[ \sum_{i=1}^{N}P_{i1}^{2}\ln \left( \frac{
m_{H_{i}^{0}}^{2}}{m_{W}^{2}}\right)
+\sum_{i=1}^{N}\sum_{j=1}^{N-1}P_{i,j+1}^{2}K\left(
m_{H_{i}^{0}}^{2},m_{A_{j}^{0}}^{2},m_{H_{j}^{\pm }}^{2}\right) \right] 
\notag \\
&=&\frac{1}{12\pi }\ln \left( \frac{m_{h}^{2}}{m_{W}^{2}}\right) +\frac{1}{%
12\pi }\left[ \sum_{i=2}^{N}P_{i1}^{2}\ln \left( \frac{m_{H_{i}^{0}}^{2}}{%
m_{h}^{2}}\right) +\sum_{i=1}^{N}\sum_{j=1}^{N-1}P_{i,j+1}^{2}K\left(
m_{H_{i}^{0}}^{2},m_{A_{j}^{0}}^{2},m_{H_{j}^{\pm }}^{2}\right) \right], 
\label{Sh}
\end{eqnarray}
where we identified the lightest CP-even Higgs $H_{1}^{0}=h$ with the LHC
Higgs-like particle with the mass $m_{h}=125$ GeV. We defined a function 
\begin{eqnarray}
K\left( m_{1}^{2},m_{2}^{2},m_{3}^{2}\right)  &=&\frac{1}{\left(
m_{2}^{2}-m_{1}^{2}\right) {}^{3}}\left\{ m_{1}^{4}\left(
3m_{2}^{2}-m_{1}^{2}\right) \ln \left( \frac{m_{1}^{2}}{m_{3}^{2}}\right)
-m_{2}^{4}\left( 3m_{1}^{2}-m_{2}^{2}\right) \ln \left( \frac{m_{2}^{2}}{%
m_{3}^{2}}\right) \right.   \notag \\
&&-\left. \frac{1}{6}\left[ 27m_{1}^{2}m_{2}^{2}\left(
m_{1}^{2}-m_{2}^{2}\right) +5\left( m_{2}^{6}-m_{1}^{6}\right) \right]
\right\} ,
\end{eqnarray}%
with the properties 
\begin{eqnarray}
\lim_{m_{1}\rightarrow m_{2}}K(m_{1}^{2},m_{2}^{2},m_{3}^{2})
&=&K_{1}(m_{2}^{2},m_{3}^{2})=\ln \left( \frac{m_{2}^{2}}{m_{3}^{2}}\right) ,
\notag \\
\lim_{m_{2}\rightarrow m_{3}}K(m_{1}^{2},m_{2}^{2},m_{3}^{2})
&=&K_{2}(m_{1}^{2},m_{3}^{2})=\frac{%
-5m_{1}^{6}+27m_{1}^{4}m_{3}^{2}-27m_{1}^{2}m_{3}^{4}+6\left(
m_{1}^{6}-3m_{1}^{4}m_{3}^{2}\right) \ln \left( \frac{m_{1}^{2}}{m_{3}^{2}}%
\right) +5m_{3}^{6}}{6\left( m_{1}^{2}-m_{3}^{2}\right) ^{3}},  \notag \\
\lim_{m_{1}\rightarrow m_{3}}K(m_{1}^{2},m_{2}^{2},m_{3}^{2})
&=&K_{2}(m_{2}^{2},m_{3}^{2}).
\end{eqnarray}

We can split the $S$ parameter as $S=S_{SM}+\Delta S$, where $S_{SM}$\ is
the contribution from the SM, while $\Delta S$ contain all the
contributions involving the heavy scalars: 
\begin{equation}
S_{SM}=\frac{1}{12\pi }\ln \left( \frac{m_{h}^{2}}{m_{W}^{2}}\right) ,
\end{equation}

\begin{equation}
\label{Delta-S-1}
\Delta S\simeq \frac{1}{12\pi }\left[ \sum_{i=2}^{N}P_{i1}^{2}\ln \left( 
\frac{m_{H_{i}^{0}}^{2}}{m_{h}^{2}}\right)
+\sum_{i=1}^{N}\sum_{j=1}^{N-1}P_{i,j+1}^{2}K\left(
m_{H_{i}^{0}}^{2},m_{A_{j}^{0}}^{2},m_{H_{j}^{\pm }}^{2}\right) \right] .
\end{equation}

\section{T and S bounds on NHDM}

\label{T-S-exp bounds}

\begin{figure}[tbh]
\centering
\includegraphics[width=10cm,height=10cm,angle=0]{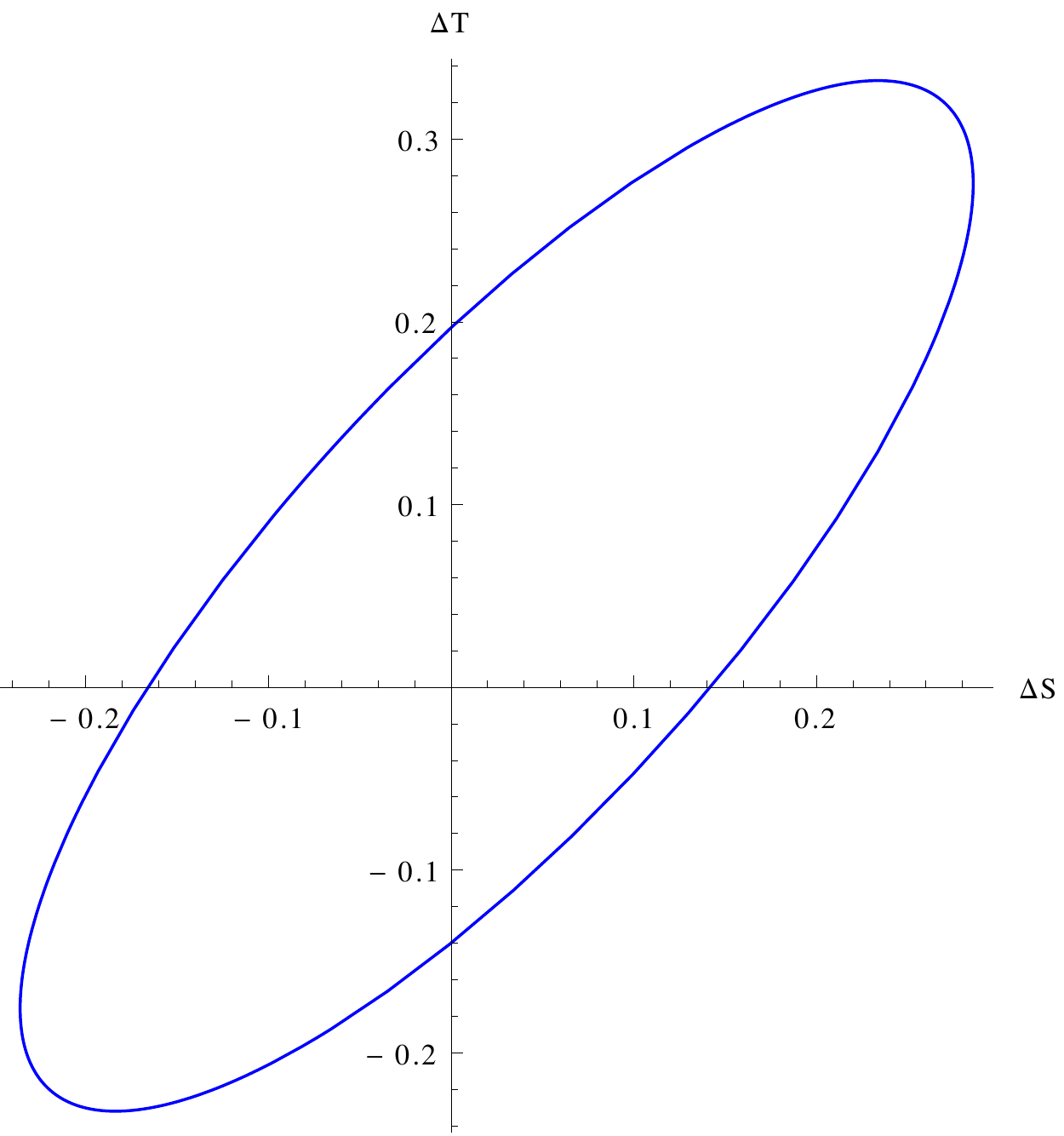} 
\caption{The interior of the ellipse in $\Delta S-\Delta T$ plane is the experimentally allowed region at 95\%C.L. from Ref. \protect\cite{Baak:2012kk}. The
reference point $\Delta S=\Delta T=0$ is conventionally taken to the Standard Model value of $\Delta S$ and $\Delta T$, at $m_h=125.7$GeV and $m_t=173.18GeV$}
\label{diag2}
\end{figure}
The inclusion of the extra scalar particles modifies the SM predictions for the oblique parameters $T$ and $S$, and therefore 
their values extracted from high 
precision measurements can be used to constrain 
the $N$ Higgs extension of the SM. Our goal is to examine if these measurements are able to restrict the number $N$ of Higgs electroweak doublets.
The experimental results on $T$ and $S$ restrict the deviations  
$\Delta T$ and $\Delta S$  from the SM predictions  to lie inside a region in 
the $\Delta S-\Delta T$
plane. At the $95\%$C.L., 
this region is the elliptic contour shown in 
Fig.~\ref{diag2}, taken from Ref.~\cite{Baak:2012kk}.  The reference point $\Delta S=\Delta T=0$ is conventionally taken to be the SM value of $\Delta S$ and $\Delta T$ at $m_h=125.7$GeV and $m_t=173.18GeV$. 
In view of the complexity of the general case of the $N$ Higgs doublet model, we consider several benchmark scenarios described below. 

\subsection{All the heavy Higgses are degenerate.}

\label{subsect: Senario-1} This is the most simple case of the Higgs
spectrum with the lightest CP-even Higgs $H_{1}^{0}=h$ identified with the LHC
Higgs-like particle, with a mass $m_{h}=125.7$ GeV and all the other heavier
Higgses degenerate having a common mass $m_{H}$. Thus, 
\begin{eqnarray}
&&m_{H_{j}^{\pm }}=m_{A_{j}^{0}}=m_{H},\hspace{4.4cm}j=1,2,\cdots ,N-1,
\label{Scenario-1} \\
\nonumber
&&m_{H_{1}^{0}}\ =m_{h}=125.7GeV,\hspace{0.5cm}m_{H_{i}^{0}}=m_{H},\hspace{%
1.5cm}i=2,\cdots ,N,  \\
\nonumber
&&m_{H}> m_{h}.
\end{eqnarray}%
For this spectrum, Eqs. (\ref{Delta-T-1}) and (\ref{Delta-S-1}) for the $\Delta T$ and $\Delta S$ parameters
are drastically simplified and take the form 
\begin{eqnarray}
\Delta T &=&-\frac{3\left( 1-P_{11}^{2}\right) }{16\pi \cos ^{2}\theta _{W}}
\ln \left( \frac{m_{H}^{2}}{m_{h}^{2}}\right) ,  \label{Delta S-S1} \\
\Delta S &=&\frac{1-P_{11}^{2}}{12\pi }\left[ \ln \left( \frac{m_{H}^{2}}
{m_{h}^{2}}\right) +K_{2}\left( m_{h}^{2},m_{H}^{2}\right) \right]. 
\end{eqnarray}

Thus, in this scenario neither of the two parameters $S$ and $T$ depends on $N$. Therefore, the spectrum in Eq. (\ref{Scenario-1}) does not constrain the
number of Higgs doublets.

\subsection{Degeneracy inside the groups of the heavy CP-even, CP-odd and charged Higgses.}
\label{subsect: Scenario-2}
{{\bf Subscenario B1:} \it The CP-even and CP-odd neutral Higgses are degenerate.}\\
The next-to-simplest scenario that we consider has the following Higgs spectrum:
\begin{eqnarray}
\label{spectrum-B1}
&&m_{A_{j}^{0}}=m_{H},\hspace{9mm}\hspace{1.5cm}m_{H_{j}^{\pm
}}=m_{H}+\Delta ,\hspace{1.5cm}\hspace{1.5cm}j=1,2,\cdots ,N-1,  \label{s2a}
\\
&&m_{H_{1}^{0}}\ =m_{h}=125.7GeV,\hspace{1.4cm}m_{H_{i}^{0}}=m_{H},\hspace{%
3.7cm}i=2,\cdots ,N,  \notag \\
&&m_{H}>m_{h}.  \notag
\end{eqnarray}
This spectrum, using  Eqs. (\ref{Delta-T-1}) and  (\ref{Delta-S-1}), leads to the expressions
\begin{figure}[tbh]
\centering
\includegraphics[width=15cm,height=10cm,angle=0]{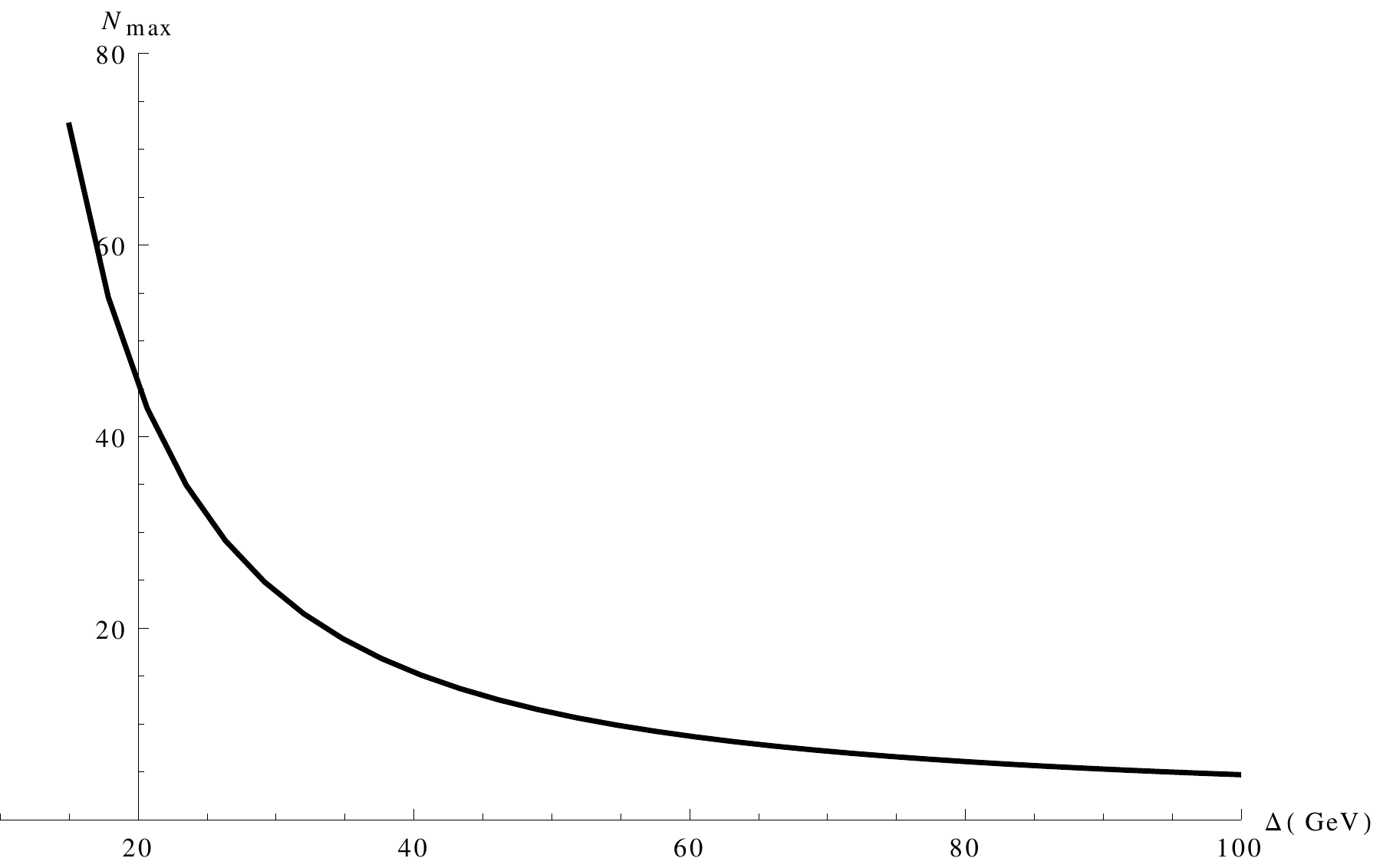} 
\caption{Upper bound $N\leq N_{max}$ on the number $N$ of Higgs doublets, obtained from $T$ and 
$S $ at $95\%$C.L., using the experimental constraints indicated in Fig.~\ref{diag2} for the Higgs spectrum in Eq.~(\ref{spectrum-B1}). }
\label{diagcaso1}
\end{figure}
\begin{eqnarray}
\label{Delta-spectrum-B1}
\Delta T &=&-\frac{3\left( 1-P_{11}^{2}\right) }{16\pi \cos ^{2}\theta _{W}}%
\ln \left( \frac{m_{H}^{2}}{m_{h}^{2}}\right) +\frac{N-1}{16\pi
^{2}v^{2}\alpha _{EM}(M_{Z})}\left[ m_{H^{\pm }}^{2}+m_{H}^{2}-2h\left(
m_{H}^{2},m_{H^{\pm }}^{2}\right) \right]  \\
&&+\frac{1-P_{11}^{2}}{16\pi ^{2}v^{2}\alpha _{EM}(M_{Z})}\left[ h\left(
m_{h}^{2},m_{H}^{2}\right) -h\left( m_{h}^{2},m_{H^{\pm }}^{2}\right)
-m_{H}^{2}+h\left( m_{H}^{2},m_{H^{\pm }}^{2}\right) \right] ,  \notag \\
\Delta S &=&\frac{1}{12\pi }\left\{ \left( 1-P_{11}^{2}\right) \left[ \ln
\left( \frac{m_{H}^{2}}{m_{h}^{2}}\right) +K\left(
m_{h}^{2},m_{A}^{2},m_{H^{\pm }}^{2}\right) -K_{1}\left( m_{H}^{2},m_{H^{\pm
}}^{2}\right) \right] +\left( N-1\right) K_{1}\left( m_{H}^{2},m_{H^{\pm
}}^{2}\right) \right\} ,  \notag
\end{eqnarray}%
with 
\begin{equation}
P_{11}=\sum_{l=1}^{N}R_{l1}Q_{l1}=\frac{1}{v}\sum_{l=1}^{N}R_{l1}v_{l},
\end{equation}
where we used the first relation in  Eq. (\ref{vev}).

Now, using Eqs.~(\ref{Delta-spectrum-B1}),  we find the maximal values $N_{max}$ of the Higgs doublets $N$ compatible with the precision data in Fig~\ref{diag2}.  We scan the parameter space within
\begin{eqnarray}\label{scan-sp-2-1}
&& 0 \leq P_{11} \leq 1, \ \ \    600 \mbox{GeV} \leq m_{H}\leq 1 \mbox{TeV}.
\end{eqnarray}
In Fig.~\ref{diagcaso1}, we show the resulting  values $N_{max}$ in a function of the splitting parameter $\Delta$.
It is noteworthy that the maximum number of Higgs doublets
decreases when the mass splitting $\Delta$ between the heavy physical scalars is increased. 
This behavior follows from the fact that increasing the number of Higgs doublets yields an increase of the $T$ and $S$ oblique parameters.

\quad In the limit $\Delta \rightarrow 0$, we find $N_{max} \rightarrow \infty$, corresponding to 
no limits on $N$, which is consistence with the scenario (\ref{Scenario-1}).  \\[3mm]
%

{{\bf Subscenario B2:} \it The CP-even neutral and charged Higgses are
degenerate:}
%
%
\begin{eqnarray}
\label{spectrum-B2}
&&m_{H_{j}^{\pm }}=m_{H},\hspace{9mm}\hspace{1.5cm}m_{A_{j}^{0}}=m_{H}+%
\Delta ,\hspace{1.5cm}\hspace{1.5cm}j=1,2,\cdots ,N-1,  \label{s2b} \\
&&m_{H_{1}^{0}}\ =m_{h}=125.7GeV,\hspace{1.4cm}m_{H_{i}^{0}}=m_{H},\hspace{%
3.7cm}i=2,\cdots ,N,  \notag \\
&&m_{H} > m_{h}.  \notag
\end{eqnarray}

\begin{figure}[tbh]
\centering
\includegraphics[width=15cm,height=10cm,angle=0]{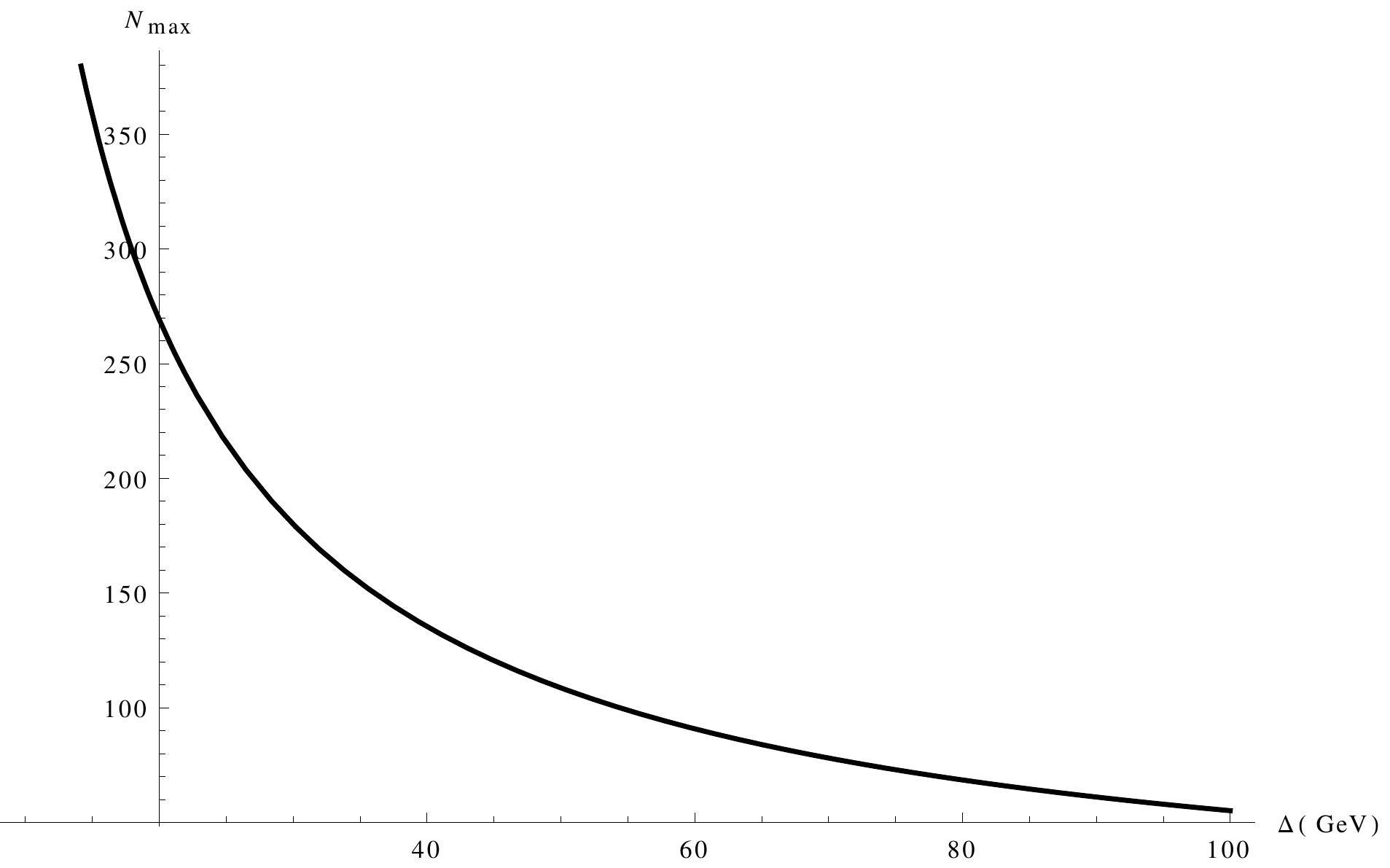} 
\caption{The same as in Fig.~\ref{diagcaso1}, but for the spectrum in 
Eq.~(\ref{spectrum-B2}). }   
\label{diagcaso2}
\end{figure}
From Eqs. (\ref{Delta-T-1}) and  (\ref{Delta-S-1}), we find for this spectrum
\begin{eqnarray}
\label{Delta-spectrum-B2}
\Delta T &=&-\frac{3\left( 1-P_{11}^{2}\right) }{16\pi \cos ^{2}\theta _{W}}%
\ln \left( \frac{m_{H}^{2}}{m_{h}^{2}}\right) +\frac{1-P_{11}^{2}}{16\pi
^{2}v^{2}\alpha _{EM}(M_{Z})}\left[ h\left( m_{h}^{2},m_{A}^{2}\right)
-h\left( m_{h}^{2},m_{H}^{2}\right) +m_{H}^{2}-h\left(
m_{H}^{2},m_{A}^{2}\right) \right] ,  \notag \\
\Delta S &=&\frac{1}{12\pi }\left\{ \left( 1-P_{11}^{2}\right) \left[ \ln
\left( \frac{m_{H}^{2}}{m_{h}^{2}}\right) +K\left(
m_{h}^{2},m_{A}^{2},m_{H}^{2}\right) \right] +\left( N-2+P_{11}^{2}\right)
K_2\left( m_{H}^{2},m_{A}^{2}\right) \right\}\label{DeltaTS}.
\end{eqnarray}
Scanning the parameter space in the region (\ref{scan-sp-2-1}), we find the maximal values on the number $N$ of  Higgs doublets compatible with the data in Fig.~\ref{diag2}. The results are shown in  Fig.~\ref{diagcaso2}. As seen, this spectrum is significantly less restrictive for 
$N$ than that in Eq. (\ref{spectrum-B1}). This is mainly because of the fact that only the $S$ parameter depends on $N$ in the present case, while in the case of  the spectrum 
(\ref{spectrum-B1}) both $T$ and $S$ are $N$-dependent.\\[3mm]
{{\bf Subscenario B3:} \it The CP-odd neutral and charged Higgses are
degenerate:}
%
\begin{eqnarray}
\label{spectrum-B3}
&&m_{H_{j}^{\pm }}=m_{H},\hspace{9mm}\hspace{1.5cm}m_{A_{j}^{0}}=m_{H},%
\hspace{1.5cm}\hspace{1.5cm}j=1,2,\cdots ,N-1,  \label{s2c} \\
&&m_{H_{1}^{0}}\ =m_{h}=125.7GeV,\hspace{1.4cm}m_{H_{i}^{0}}=m_{H}+\Delta ,%
\hspace{3.7cm}i=2,\cdots ,N,  \notag \\
&&m_{H} > m_{h}.\notag
\end{eqnarray}
In this case we find from Eqs. (\ref{Delta-T-1}) and (\ref{Delta-S-1})
%
\begin{eqnarray}
\label{Delta-spectrum-B3}
\Delta T &\simeq &-\frac{3\left( 1-P_{11}^{2}\right) }{16\pi \cos ^{2}\theta
_{W}}\ln \left( \frac{m_{H^{0}}^{2}}{m_{h}^{2}}\right)   \notag \\
\Delta S &\simeq &\frac{1}{12\pi }\left\{ \left( 1-P_{11}^{2}\right) \left[
\ln \left( \frac{m_{H^{0}}^{2}}{m_{h}^{2}}\right) +K_2\left(
m_{h}^{2},m_{H}^{2}\right) \right] +\left( N-2+P_{11}^{2}\right)
K_2\left( m_{H^{0}}^{2},m_{H}^{2}\right) \right\} .
\end{eqnarray}
\begin{figure}[tbh]
\centering
\includegraphics[width=15cm,height=10cm,angle=0]{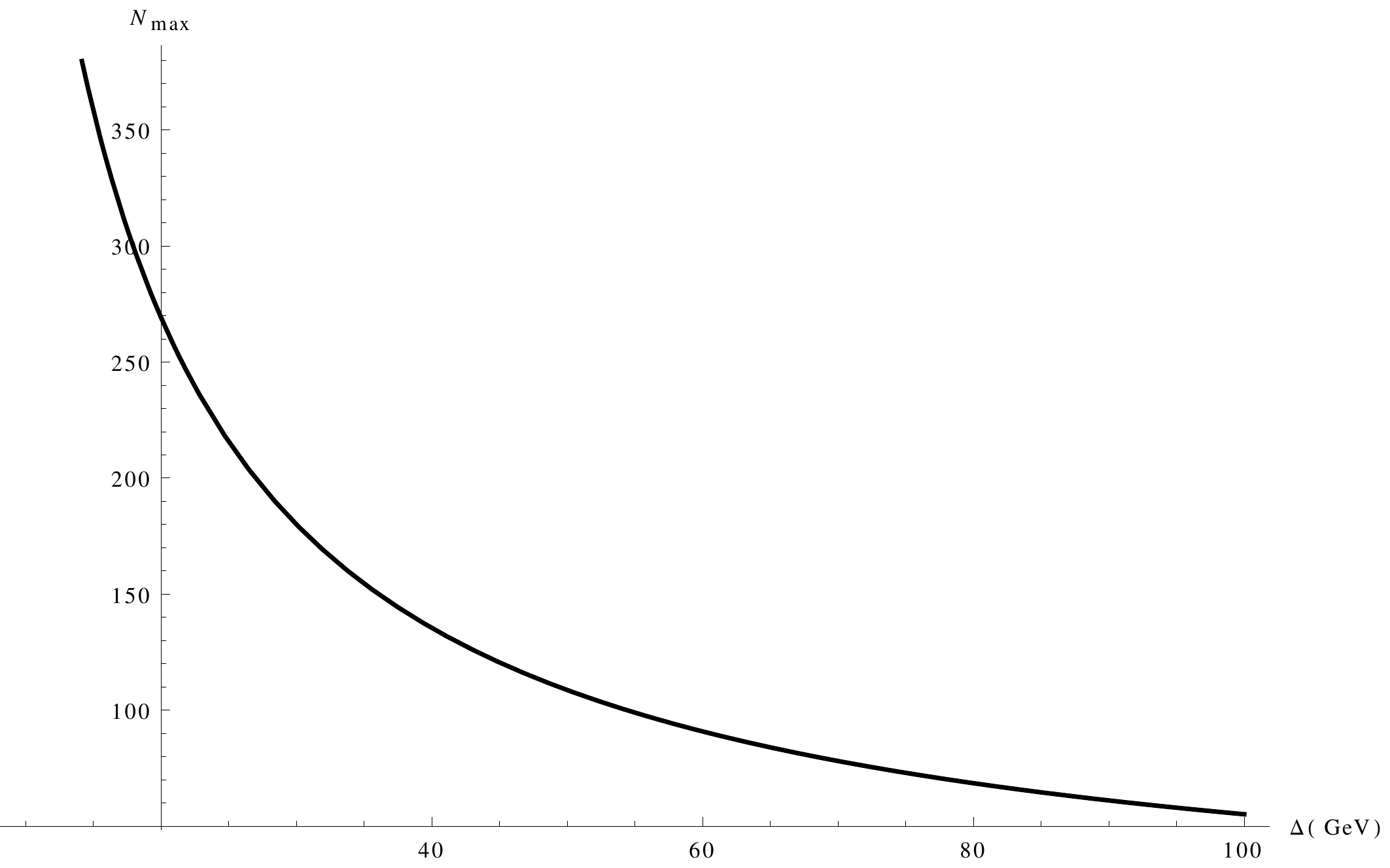} 
\caption{The same as in Fig.~\ref{diagcaso1}, but for the spectrum in 
Eq.~(\ref{spectrum-B3}). }   
\label{diagcaso3}
\end{figure}
Applying the same procedure as previously, we scan the parameter space in the region (\ref{scan-sp-2-1}) and find the maximal values on the number $N$ of the Higgs doublets compatible with the data of Fig.~\ref{diag2}. The results are shown in Fig.~\ref{diagcaso3}. 
Again, as in the case (\ref{spectrum-B2}), the parameter $T$ is independent of $N$. 
As a consequence, the limits on $N$ for the spectrum (\ref{spectrum-B3}) are significantly weaker than for (\ref{spectrum-B1}).\\[3mm]

\begin{figure}[tbh]
\centering
\includegraphics[width=9cm,height=9cm,angle=0]{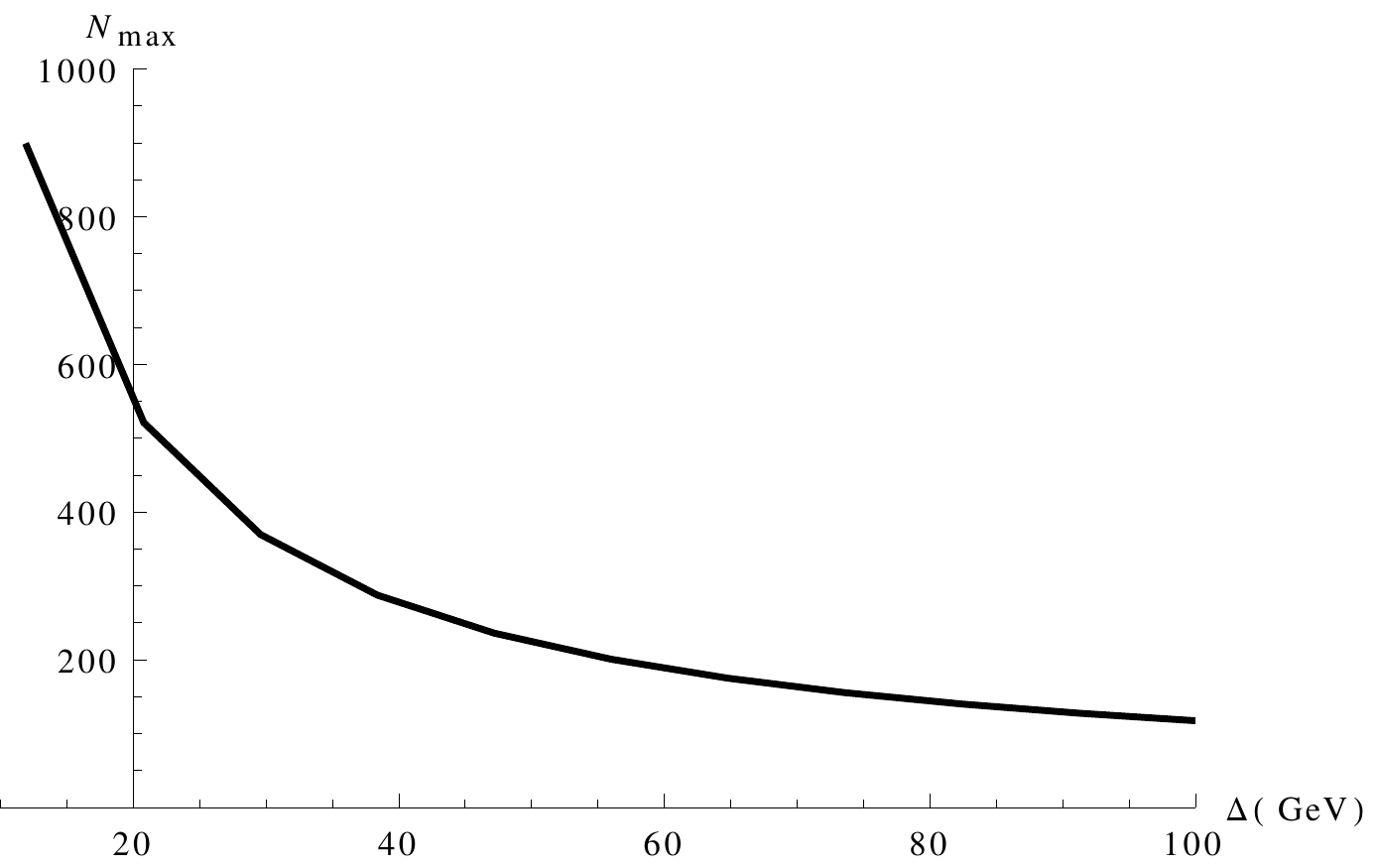}%
\includegraphics[width=9cm,height=9cm,angle=0]{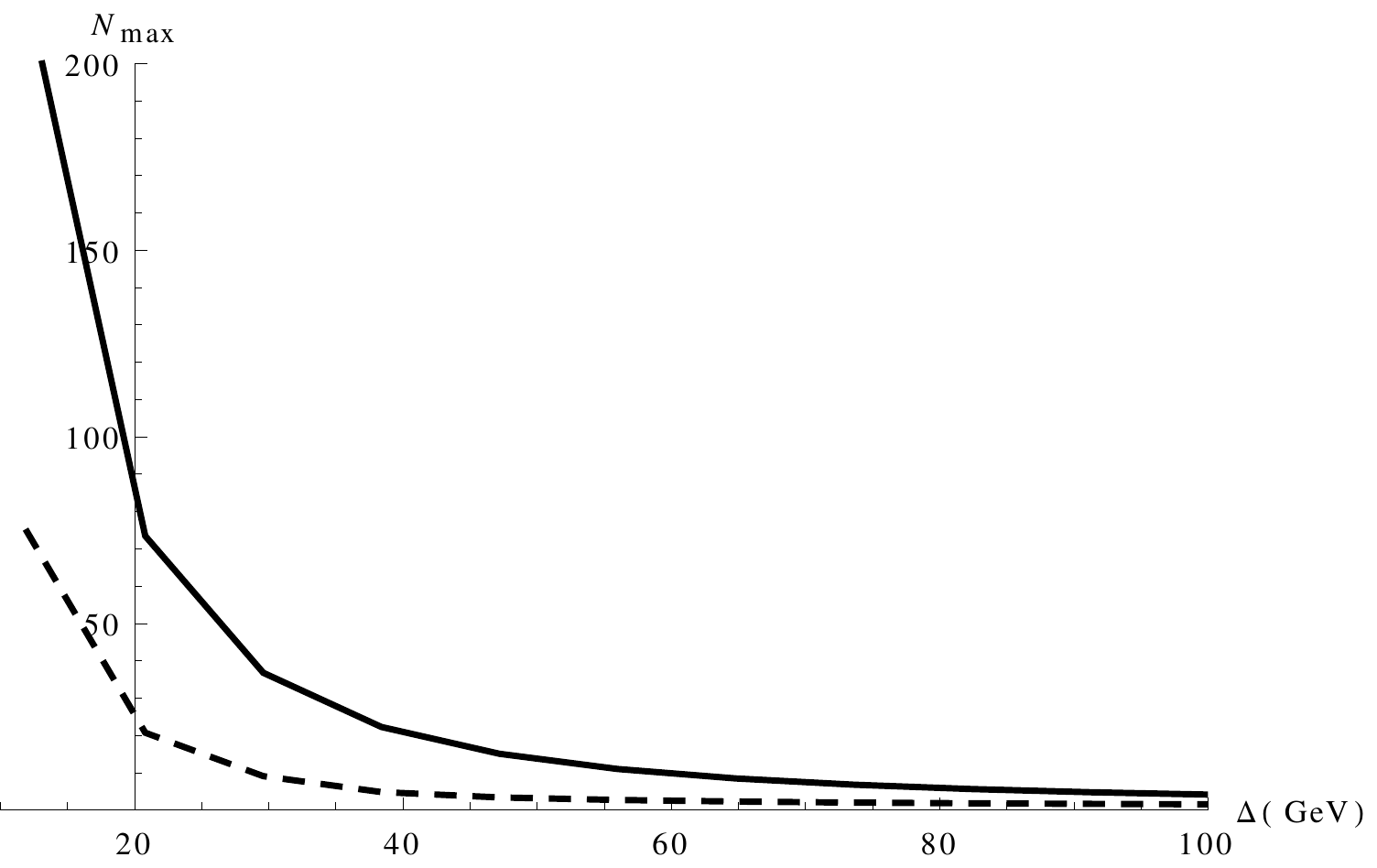} 
\hspace{2cm}{\footnotesize {$a=1$}}\hspace{9cm}{\footnotesize {$a=0.5,2$}}%
\hspace{0cm}\newline
\caption{The same as in Fig.~\ref{diagcaso1}, but for the spectrum in 
Eq.~(\ref{spectrum-B4}). The left panel is for $a = 1$; the right panel is for $a = 0.5$ (dashed line), $a = 2$ (solid line).  }   
\label{diagpdg1}
\end{figure}

\begin{figure}[tbh]
\centering
\includegraphics[width=9cm,height=9cm,angle=0]{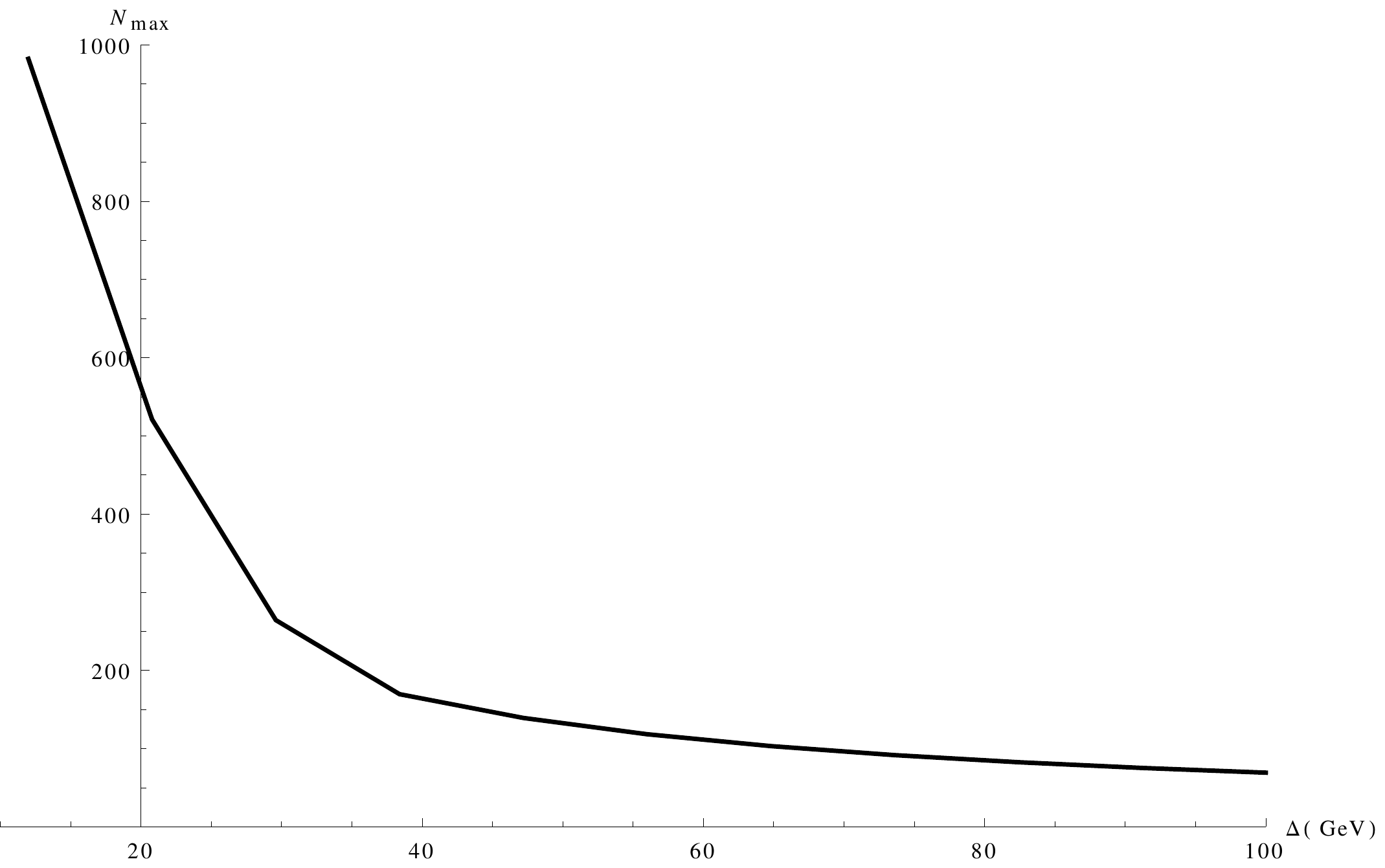}%
\includegraphics[width=9cm,height=9cm,angle=0]{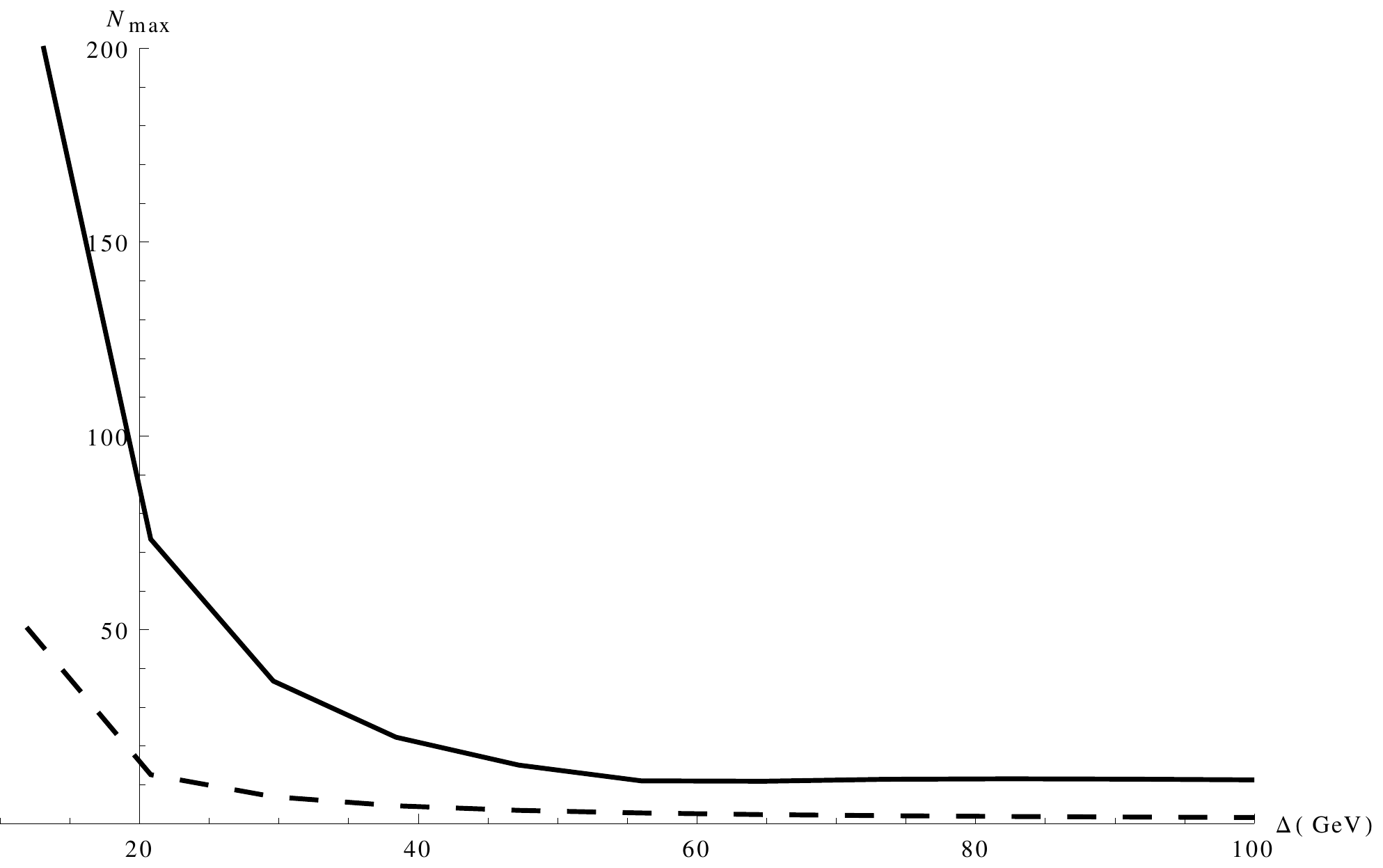} 
{\footnotesize {$a=1$}}\hspace{9cm}{\footnotesize {$a=0.5,2$}}\hspace{0cm}%
\caption{The same as in Fig.~\ref{diagpdg1}, but for the spectrum in 
Eq.~(\ref{spectrum-B5}). 
}
\label{diagpdg2}
\end{figure}

\begin{figure}[tbh]
\centering
\includegraphics[width=15cm,height=10cm,angle=0]{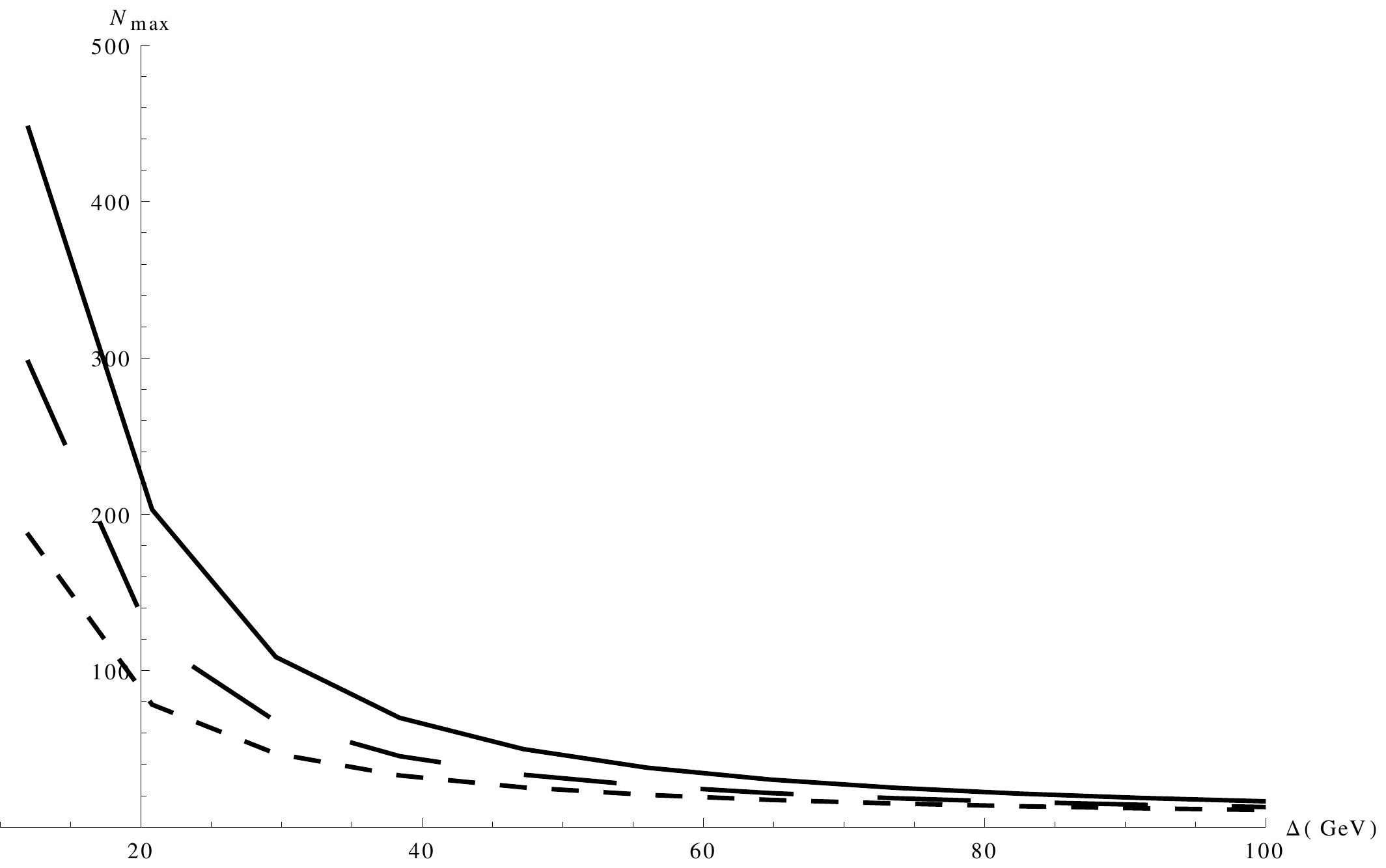} 
\caption{The same as in Fig.~\ref{diagcaso1}, but for the spectrum in 
Eq.~(\ref{spectrum-B6}). The curves from the bottom to the top
correspond to $a =0.5,1,2$, respectively. }
\label{diagpdg3}
\end{figure}

{{\bf Subscenario B4:} \it Split groups with the degenerate interior:}
\begin{eqnarray}
&\mbox{Case 1}: & m_{A_{j}^{0}}=m_{H}+\Delta ,\hspace{9mm}\hspace{1.5cm}
m_{H_{j}^{\pm}}=m_{H}+a\Delta ,\hspace{1.5cm} j=1,2,\cdots ,N-1,  \notag \\
&&m_{H_{1}^{0}}\ =m_{h}=125.7GeV,\hspace{1.4cm}m_{H_{i}^{0}}=m_{H},
\hspace{3.7cm}i=2,\cdots ,N,  \notag \\
&&m_{H} > m_{h}.  \label{spectrum-B4}\\[3mm]
&\mbox{Case 2}: & m_{H_{i}^{0}} =m_{H}+\Delta ,\hspace{9mm}\hspace{1.5cm}m_{H_{j}^{\pm}}=m_{H}+a\Delta ,\hspace{1.5cm} j=1,2,\cdots ,N-1,  \notag \\
&&m_{H_{1}^{0}}\ =m_{h}=125.7GeV,\hspace{1.4cm}m_{A_{j}^{0}}=m_{H},
\hspace{3.7cm}i=2,\cdots ,N,  \notag \\
&&m_{H} > m_{h}.  \label{spectrum-B5}\\[3mm]
&\mbox{Case 3}: & m_{H_{i}^{0}} =m_{H}+\Delta ,\hspace{9mm}\hspace{1.5cm}
m_{A_{j}^{0}}=m_{H}+a\Delta ,\hspace{1.5cm}
j=1,2,\cdots ,N-1,  \notag \\
&&m_{H_{1}^{0}}\ =m_{h}=125.7GeV,\hspace{1.4cm}m_{H_{j}^{\pm }}=m_{H},
\hspace{3.7cm}i=2,\cdots ,N,  \notag \\
&&m_{H} > m_{h}.  \label{spectrum-B6}
\end{eqnarray}
For all these cases we find from Eqs. (\ref{Delta-T-1}) and  (\ref{Delta-S-1})
\begin{eqnarray}
\label{Delta-spectrum-B4}
\Delta T &=&-\frac{3\left( 1-P_{11}^{2}\right) }{16\pi \cos ^{2}\theta _{W}}%
\ln \left( \frac{m_{H^{0}}^{2}}{m_{h}^{2}}\right) +  \notag \\
&&+\frac{1-P_{11}^{2}}{16\pi ^{2}v^{2}\alpha _{EM}(M_{Z})}\left[ h\left(
m_{H^{0}}^{2},m_{H^{\pm }}^{2}\right) -h\left(
m_{H^{0}}^{2},m_{A}^{2}\right) +h\left( m_{h}^{2},m_{A}^{2}\right) -h\left(
m_{h}^{2},m_{H^{\pm }}^{2}\right) \right] +  \notag \\
&&+\frac{N-1}{16\pi ^{2}v^{2}\alpha _{EM}(M_{Z})}\left[ m_{H^{\pm
}}^{2}-h\left( m_{A}^{2},m_{H^{\pm }}^{2}\right) +h\left(
m_{H^{0}}^{2},m_{A}^{2}\right) -h\left( m_{H^{0}}^{2},m_{H^{\pm
}}^{2}\right) \right] ,  \notag \\[0.12in]
\Delta S &=&\frac{1}{12\pi }\left\{ \left( 1-P_{11}^{2}\right) \left[ \ln
\left( \frac{m_{H^{0}}^{2}}{m_{h}^{2}}\right) +K\left(
m_{h}^{2},m_{A}^{2},m_{H^{\pm }}^{2}\right) \right] +\left(
N-2+P_{11}^{2}\right) K\left( m_{H^{0}}^{2},m_{A}^{2},m_{H^{\pm
}}^{2}\right) \right\}.
\end{eqnarray}
As seen from Eq. (\ref{Delta-spectrum-B4}), now both parameters $T$ and $S$ depend
linearly on $N$, the number of Higgs doublets. 
Scanning the parameter space in the region (\ref{scan-sp-2-1}), we find the results for several sample values of the parameter $a = 0.5, 1, 2$  shown in Figs.~\ref{diagpdg1}-\ref{diagpdg3}.
Note the general tendency:  the large splitting, corresponding to the larger values of $\Delta$ and 
$a$, leads to more stringent constraints on $N$.  

\label{Scenario-3}

\begin{figure}[tbh]
\centering
\hspace{-20mm}%
\includegraphics[width=15cm,height=10cm,angle=0]{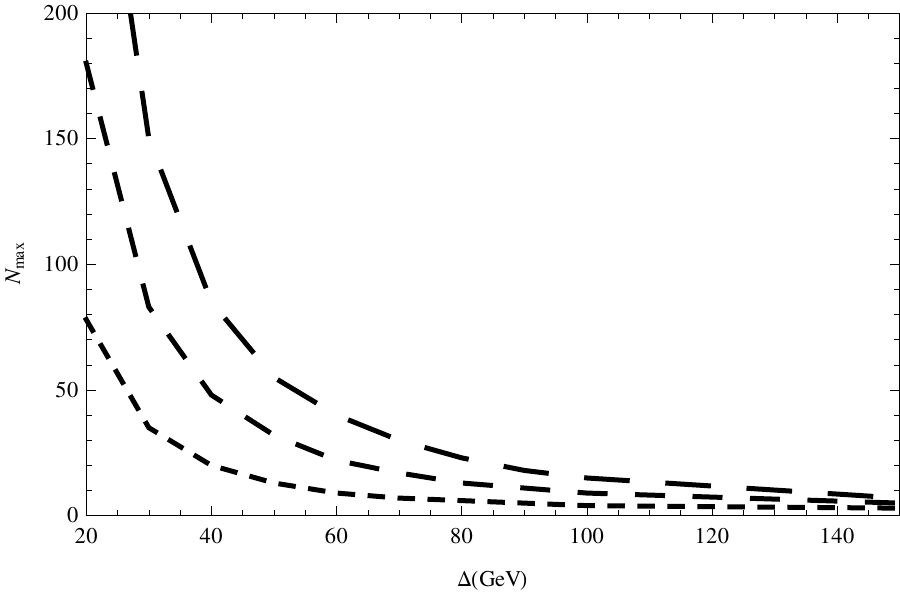} %
\caption{The same as in Fig.~\ref{diagcaso1}, but for the spectrum Eq.~(\ref{No-Deg}).
The curves from the bottom to the top correspond to $n= 20,10,2$,
respectively.}
\label{diagnodeg}
\end{figure}

\subsection{No degeneracy with a particular structure of the spectrum}
Finally, let us consider a benchmark scenario in which all the Higgses are
nondegenerate. Since the general case can hardly be analyzed, we consider a particular structure of the Higgs spectrum: 
%
\begin{eqnarray}
\label{spectrum-C}
m_{H_{1}^{0}} &=&m_{h}=125.7GeV,\hspace{1.5cm}\hspace{3mm}m_{H_{i}^{0}}\
=m_{H}+2\left( i-2\right) \Delta ,\hspace{1.5cm}i=2,\cdots ,N,
\label{No-Deg} \\
\nonumber
m_{A_{j}^{0}} &=&m_{H}+\left( 2j-1\right) \Delta ,\hspace{1.5cm}%
m_{H_{j}^{\pm }}=m_{H}+2\left( j-1\right) \Delta +\delta ,\hspace{8mm}%
j=1,2,\cdots ,N-1,\\
\nonumber
m_{H} &>& m_{h}.
\end{eqnarray}
This is a spectrum equidistant in each of  the three groups of Higgses 
$\{A^{0}_{j}\}, \{H^{0}_{i}\}, \{H^{\pm}_{j}\}$ with a step $2\Delta$. The bands of these groups overlap with each other. The spectrum is characterized by two parameters $\Delta$ 
and $\delta$. 

For simplicity, we assume 
\begin{equation}
P_{i,j}=\delta _{ij},\hspace{1.5cm}\hspace{1.5cm}i,j=i=1,2,\cdots ,N
\end{equation}
Then from Eqs. (\ref{Delta-T-1}) and  (\ref{Delta-S-1}) we get
\begin{eqnarray}
\Delta T &\simeq &\frac{1}{16\pi ^{2}v^{2}\alpha _{EM}(M_{Z})}%
\sum_{i=1}^{N-1}\left[ m_{H_{i}^{\pm }}^{2}-h\left(
m_{A_{i}^{0}}^{2},m_{H_{i}^{\pm }}^{2}\right) +h\left(
m_{H_{i+1}^{0}}^{2},m_{A_{i}^{0}}^{2}\right) -h\left(
m_{H_{i+1}^{0}}^{2},m_{H_{i}^{\pm }}^{2}\right) \right] ,  \label{Delta S-S3}
\\
\Delta S &\simeq &\frac{1}{12\pi }\sum_{i=1}^{N-1}K\left(
m_{H_{i+1}^{0}}^{2},m_{A_{i}^{0}}^{2},m_{H_{i}^{\pm }}^{2}\right) .
\end{eqnarray}
Scanning the parameter space in the region (\ref{scan-sp-2-1}), we find the results for several sample values of the parameter $\delta = \Delta/n$ 
shown in Fig.~\ref{diagnodeg}. 
The curves from the bottom to the top correspond to 
$n=20,10,2$, respectively.  
With the larger value $n=50$, we find 
$N_{max} \sim 570$ for $\Delta =20$ GeV. 

\section{Summary and Conclusions}
\label{Conclusions}
We have considered an $N$ Higgs $SU(2)$ doublet model (NHDM) with arbitrary number 
$N$. In this model, we calculated the one-loop contributions $\Delta S$ and $\Delta T$ of the Higgs doublets to the electroweak oblique parameters $S$ and $T$. The calculated contribution depends on the number $N$ of Higgs doublets, and, therefore, our results can be used to constrain $N$ from data on the precision measurements of the parameters $T$ and $S$. 
Within the generic case of the NHDM, due to the large number of free parameters, this program can hardly be realized. For this reason, we have analyzed several benchmark scenarios with particular mass spectra [Eqs.~(\ref{Scenario-1}), (\ref{spectrum-B1}), (\ref{spectrum-B2}), (\ref{spectrum-B3}), (\ref{spectrum-B4})-(\ref{spectrum-B6}), and (\ref{No-Deg})] of the physical scalars of the NHDM, including some other simplifying assumptions, inspired by the well-known case of the 2HDM, about  
the physical Higgs mixing and the vacuum structure of the model. These scenarios correspond to certain domains of the NHDM parameter space.  We have shown that, except for a very particular ``fine-tuned'' case with all the physical heavy Higgses degenerate [Eq.~(\ref{Scenario-1})], these scenarios imply constraints on the number of Higgs doublets $N$, in order to be compatible with the existing data on the precision measurements of $T$ and $S$.

We presented our results on $N\leq N_{max}$  
in Figs.~\ref{diagcaso1}-\ref{diagnodeg} as functions of the mass splitting parameter $\Delta$.
The general feature of our results is that the maximal number $N_{max}$ of  Higgs doublets is a monotonically increasing function for small values of the splitting 
\mbox{$\Delta \leq 20$ GeV} and a monotonically decreasing  one for larger values. Thus, the data on $T$ and $S$ are able to accommodate an arbitrary large number $N$ with decreasing splitting between the masses of the physical scalars, and, vice versa, $N$ becomes stringently constrained in those parts of the NHDM parameter space with large mass splitting in the scalar sector. The same tendency is demonstrated by the plots with respect to their dependence on the two other additional parameters $a$ and $\delta$, characterizing the Higgs mass spectrum: a smaller mass splitting corresponds to a larger number $N$ compatible with the analyzed data and visa versa. This is the main message of the present study. Also worth mentioning is that the maximal number $N_{max}$ of Higgs doublets is exactly the same when the charged Higgses are degenerate between either the CP-even or the CP-odd neutral Higgses, as shown in Figs.~\ref{diagcaso2},\ref{diagcaso3}, and larger than the obtained in the scenario of CP-odd - CP-even neutral Higgses degeneracy. Consequently, the tightest $T$ and $S$ oblique parameter constraints arise in the scenario where the charged Higgses are split in mass between either of the neutral CP-even or CP-odd Higgses, as shown in Fig.~\ref{diagcaso1}. This indicates that making the charged Higgses degenerate between either the neutral CP-even or the neutral CP odd Higgses helps avoid constraints from electroweak precision observables, a feature already present in the 2HDM.

\quad Our analysis cannot exclude a deviation from this tendency in certain parts of the NHDM parameter space; however, in our opinion, this should be related with certain ``fine-tuning'' of the parameters, as in the case of the spectrum (\ref{Scenario-1}). Naturally, the concrete limit on $N$ depends on a particular scenario within the generic NHDM framework. We hope our results will help examine such scenarios regarding their consistence with the present and future data on the precision measurements of the electroweak oblique parameters $T$ and $S$.

\quad As a final remark, let us note that the maximum number of Higgs multiplets is constrained from the requirement of perturbative unitarity in the scattering amplitudes of two transversely polarized $W$ bosons into a scalar pair, as shown in detail in Ref. \cite{Hally:2012pu}. From the results given in Ref. \cite{Hally:2012pu}, it follows that the maximum number of Higgs doublets consistent with the aforementioned perturbative unitarity requirement is 2307, which is larger than our obtained upper bounds on the number of Higgs doublets, from oblique $T$ and $S$ parameter constraints, for the several benchmark scenarios considered in this paper. Consequently, our 
upper bounds on the number of Higgs doublets  
are consistent with the 
perturbative unitarity in the scattering amplitudes of two transversely polarized $W$ bosons into a scalar pair.


\begin{acknowledgments}
This work has been partially supported by \mbox{FONDECYT} Projects No. 11130115, No. 1150792 and No. 110287, Centro-Cient\'{\i}fico-Tecnol\'{o}gico de Valpara\'{\i}so and DGIP internal Grant No. 111458. We thank an anonymous referee for very valuable comments.
\end{acknowledgments}

\end{document}